\documentclass[aps,pre,onecolumn,10pt]{revtex4-2}
\usepackage{amsmath}
\usepackage{amssymb}
\usepackage{amsthm}
\usepackage{bm}
\usepackage[shortlabels]{enumitem}
\usepackage{tabularx}
\usepackage{graphicx}
\usepackage{tikz-cd}
\usepackage{xcolor}
\usepackage{hyperref}
\usepackage[normalem]{ulem}
\usepackage{comment}

\usepackage[english]{babel}
\usepackage[letterpaper,top=2cm,bottom=2cm,left=3cm,right=3cm,marginparwidth=1.75cm]{geometry}

\begin{document}

\title{Extending the Duchon-Robert framework for anomalous dissipation to compressible fluid flows}
\author{Georgy Zinchenko and J\"org Schumacher}
\affiliation{Institute of Thermodynamics and Fluid Mechanics, Technische Universität Ilmenau, P.O.Box 100565, D-98684 Ilmenau, Germany}
\date{\today}

\begin{abstract}
Anomalous dissipation, the persistence of a finite mean kinetic energy dissipation as the Reynolds number tends to infinity, occurs in flows with sufficiently spatially rough velocity fields. Compressible turbulence adds further anomalous dissipation mechanisms, which we investigate in this work. To this end, the Duchon–Robert framework (DR) for anomalous dissipation is extended from the incompressible to the compressible Navier–Stokes flow case. We obtain three integral dissipation terms, two anomalous and a viscous one, which arise from the pressure–dilatation and density variations, differently from the incompressible case. Subsequently, fully compressible one-dimensional flows with traveling and mutually crossing shock waves are analysed in detail. In such flows, DR reveals a local maximum of anomalous dissipation at the shock front. Furthermore, DR is compared with a coarse-grain cascade theory of compressible turbulence due to Aluie (AL) and the relevant dissipation flux terms of both frameworks are identified and compared with each other. The comparison shows that each contribution related to the compressibility effects in DR  has its analogue in AL. Finally, a piecewise linear shock-type velocity profile, which approximates the crossing of two shock waves from the simulations, is used for an analytical analysis of the anomalous dissipation terms of DR to analyse the dependence of the terms on the local Hölder exponent. Our work is a first step towards a comparison of coherent flow structures in a compressible turbulent flow and related anomalous dissipation.
\end{abstract}

\maketitle

\section{Introduction}
In a fully developed three-dimensional compressible Navier–Stokes flow $\bm u({\bm x},t)$, the kinetic energy of fluid motion is dissipated at smaller scales in multiple ways \cite{Lele1994,Landau1987}. A fraction is directly converted into heat by the finite molecular dynamic viscosity $\mu$ of the fluid in the viscous subrange, the latter of which starts at a typical mean viscous scale, such as the Kolmogorov length \cite{Frisch1995}. However, unlike in incompressible turbulence, density and pressure fluctuations introduce additional mechanisms for energy dissipation. These genuine effects due to compressibility can change the kinetic energy transfer in the inertial subrange, especially in regimes with Mach numbers $M\gtrsim 1$, when the characteristic velocity $U$ of the problem at hand is larger than the speed of sound $a_s$. As a result, the dissipation of kinetic energy in compressible flows is regulated not only by viscosity $\mu$, but also by complex interactions of shock waves and significant gradients of pressure, density, and velocity \cite{Lee_Lele_Moin_1993}.

The concept of {\em anomalous dissipation} or {\em dissipative anomaly}, loosely speaking, a dissipation without molecular viscosity, originates from the incompressible turbulence literature \cite{Frisch1995,Dubrulle2019}. The mean rate of kinetic dissipation has been found to be finite even as the viscosity $\mu$ goes to zero.  Onsager's conjecture \cite{Onsager1949}, made in 1949, states that in the limit of zero viscosity, weak solutions to the Euler equation, namely those with “sufficiently rough” velocity fields, should have a non-zero energy loss rate even when viscosity is zero. Empirical evidence for the existence of a dissipative anomaly has been collected from experimental and numerical simulation data in refs. \cite{Sreenivasan1984,Sreenivasan1998,Kaneda2003}, see also \cite{Sreenivasan2025}.  These are known as "dissipative Euler solutions" \cite{Laszlo2014}. In other words, the dissipative anomaly is the statement that if the spatial irregularity or roughness of the velocity field is strong enough in the inertial subrange, the energy continues to cascade downwards to smaller scales and be dissipated when $\mu\to 0$. A characteristic of spatial roughness is the Hölder exponent $h$, which is determined through the velocity increments over a spatial distance $\xi=|{\bm \xi}|$, i.e., $|{\bm u}({\bm x}+{\bm \xi})-{\bm u}({\bm x})|\sim |{\bm \xi}|^h$. If $h<1/3$, the flow is rough enough to maintain the existence of anomalous dissipation, which was proven by Isett for the Euler case \cite{Isett2018}. Conversely, for a spatially smooth field (as in the viscous subrange) we have a Hölder exponent of $h=1$. 

In a recent work \cite{Zinchenko2024}, we analysed precursors of anomalous dissipation by means of a framework due to Duchon and Robert \cite{Duchon2000}, termed Duchon-Robert framework (DR) in the following, for three-dimensional incompressible Navier-Stokes turbulence. This framework starts with a local kinetic energy balance that is obtained by combining the velocity field ${\bm u}$ and the corresponding velocity field ${\bm u}^{\varepsilon}$, which is regularized over a filter or regularization scale $\varepsilon$. In the limit of $\varepsilon\to 0$, an additional anomalous dissipation term $D({\bm u})$ in the balance follows. Our intention in \cite{Zinchenko2024} was to connect characteristic coherent features of incompressible turbulence, such as stretched vortices obtained in kinematic models \cite{Burgers1948,Pullin1998,Hatakeyama1997,Kambe2000} and direct numerical  simulations \cite{Schumacher2007,Hamlington2008,Buaria2020,Pushenko2024}, to anomalous dissipation.

The present work focuses on two major points. First, we extend the DR framework for the kinetic energy balance to compressible turbulence. This extension results in two additional terms next to $D({\bm u})$ that will contribute to anomalous dissipation. We will subsequently compare these terms with an alternative formulation based on density-weighted coarse graining, known as Favre averaging \cite{Favre1969}, which was introduced by Aluie \cite{Aluie2013} and subsequently discussed in detail in refs. \cite{Eyink2018,Aluie2019}. We will term this framework the Aluie framework (AL) in the following. Secondly, the anomalous terms which arise in both frameworks are compared to each other for two simple one-dimensional compressible flows, a fully compressible case with a balance equation for the internal energy and the corresponding configuration for which the energy balance is reduced to an adiabatic equation of state. We analyse the physical processes that are responsible for this kind of dissipation in compressible flows. Therefore, we study here one-dimensional problems from gas dynamics. Particularly, we highlight the birth of shock waves and describe how they provide an effective way to transfer energy through scales. Indeed, theoretical analysis by Eyink and Drivas \cite{Eyink2018} has found an explicit “pressure-work” channel of anomalous dissipation, by which work, that is done by the compression of fluid elements, is necessarily transformed into internal energy. We relate this process to the real dynamics of shocks: incipient shock waves become steeper with increasing local Mach number, significantly increasing anomalous dissipation. To this end, successively coarse-grained analysis and analytical modeling are assisted by direct numerical simulations (DNS) of nonlinear problems, which cause the shock discontinuities. By performing a spatial filtering operation using wavelet theory \cite{Mallat1999}, we analyse energy flow at various scales and calculate the anomalous dissipation directly. Finally, we discuss anomalous dissipation in a simple analytical model, which approximates the velocity profile during the crossing of shocks as piecewise linear, to obtain closed expressions of the anomalous dissipation terms in dependence on the local Hölder regularity.

The modeling helps us identify the local flow characteristics that act as precursors to anomalous dissipation, such as the emergence of steep gradient areas or colliding waves indicating the emergence of a shock. The detection of such precursors in the velocity, density, and pressure fields helps us to attribute the emergence of anomalous dissipation to the evident real-time flows that are observed. We also calculate the anomalous dissipation terms through DR and AL to qualitatively validate the similarity of these methods.

The outline of the manuscript is as follows. In Sec. \ref{Sec::2}, we construct the Duchon-Robert framework for fully compressible Navier-Stokes flows by deriving the local energy balance. Section \ref{Sec::3} introduces the Aluie framework in brief and formulates the corresponding energy budget. In Sec. \ref{Sec::4}, we compare DR and AL, identifying their similar structure. Section \ref{Sec::5} presents our numerical implementation of fully compressible one-dimensional flow, the solution of which is analysed using both frameworks. We quantify the anomalous kinetic energy dissipation in Sec. \ref{Sec::6}. In Sec. \ref{Sec::5b}, we compare fully compressible flow with flow under adiabatic conditions. Section \ref{Sec::7} provides the analytical investigation of anomalous dissipation in a one-dimensional configuration of crossing shock waves. We finish the work with a summary and an outlook in Sec. \ref{Sec::8}

\section{Framework by Duchon and Robert \cite{Duchon2000}} \label{Sec::2}

In the study of turbulent flows, DR provides one possibility to analyse energy dissipation. Originally developed for incompressible flows \cite{Duchon2000}, this approach is now extended to compressible turbulence to account for density variations as well as pre-shocks and shocks. Here, we adapt this framework to the compressible Navier-Stokes (NS) equations, aiming to derive a local kinetic energy balance that will include anomalous dissipation terms arising from nonlinear interactions between velocity, density, and pressure fields. We begin with the compressible three-dimensional NS equations written in terms of the density-weighted velocity field,
\begin{equation}
{\bm w}({\bm x},t)=\sqrt{\rho({\bm x},t)}\, {\bm u}({\bm x},t)\,,
\end{equation}
where $\rho({\bm x},t)$ is the mass density field and ${\bm u}=(u_x,u_y,u_z)$ is the velocity field. For the rest of the work, we switch to index notation for vector and tensor fields and apply the Einstein summation rule. Indices $i,j,k=1,2,3$ or $x,y,z$ for the three-dimensional case. This choice of density-weighted velocity was made because the square of it is equal to the local kinetic energy, $E(x_j,t)=w_i^2(x_j,t)/2$ \cite{Kida1990}.  In index form, the momentum balance or Navier-Stokes equation is given by
\begin{equation}
 \begin{aligned}
    \partial_t w_i + \partial_j (u_j w_i) - \frac{w_i \theta}{2} &= \frac{1}{\sqrt{\rho}} \partial_j \tau_{ij} - \frac{1}{\sqrt\rho} \partial_i p, \\
    \theta &= \partial_j u_j,  \\
    \tau_{ij} &= \mu \left( \partial_j u_i + \partial_i u_j - \frac{2}{3} \delta_{ij} \partial_k u_k \right).
    \label{eq:1_dens_w_vel}
\end{aligned}   
\end{equation}
Here, $p$ is the pressure field and $\tau$ is the viscous stress tensor and $\theta$ is the velocity divergence. Furthermore, the second dynamic or bulk viscosity is set to zero in the strain--stress relation. The variables $u_j$ and $w_j$ denote the $j$-th components of the velocity fields ${\bm u}$ and ${\bm w}$ respectively, and $\partial_j$ is the partial derivative with respect to the $j$-th spatial coordinate $x_j$. It is also important to note that, following the DR framework, we do not use the energy equation and will focus only on the kinetic energy balance.

To apply the DR to compressible flow, the original equations \eqref{eq:1_dens_w_vel} have to be rewritten in an alternative form that isolates terms involving density-pressure and density-velocity coupling,
\begin{equation}
   \begin{aligned}
   \partial_t w_i + \partial_j (u_j w_i) - \frac{w_i \theta}{2} &= -\partial_i (\chi p) + \partial_j (\chi \tau_{ij}) + p \lambda_i - \tau_{ij} \lambda_j,  \\ 
   \chi& = \frac{1}{\sqrt{\rho}}, \\
  \quad \lambda_i &= \partial_i \left(\frac{1}{\sqrt{\rho}}\right). 
  \label{eq:2_dens_w_vel_DR}
\end{aligned} 
\end{equation}
This reformulation facilitates the identification of terms contributing to energy flux and dissipation when convolution with a test function is performed, as detailed next.

The Duchon-Robert framework employs spatial filtering to analyse energy transfer across scales. Let $\varphi\in C^{\infty}(V)$ be an infinitely differentiable function with compact support in the flow volume $V$ (which will be a 3-torus for simplicity) and normalized so that $\int_V \varphi dV=1$. We then define the rescaled function for the spatial coordinates ${\bm \xi}=(\xi_x,\xi_y,\xi_z)$ as
\begin{equation}
    \varphi^\varepsilon({\bm \xi})=\frac{1}{\varepsilon^3}\,\varphi\left(\frac{{\bm \xi}}{\varepsilon}\right)\,,
    \label{eq:3_test_func}
\end{equation}
where $\varepsilon>0$ serves as the filtering or coarse-grain scale. By denoting the convolutions ${w_i}^\varepsilon = \varphi^\varepsilon * {w_i}$, $(u_j w_i)^\varepsilon = \varphi^\varepsilon * (u_j w_i)$, $(w_i \theta)^\varepsilon = \varphi^\varepsilon * ( w_i \theta)$ as well as $(\chi p)^\varepsilon = \varphi^\varepsilon * (\chi p)$,  $(\chi \tau_{ij})^\varepsilon = \varphi^\varepsilon * (\chi \tau_{ij})$, $(p \lambda_i)^\varepsilon = \varphi^\varepsilon * (p \lambda_i)$, and $( \tau_{ij}\lambda_j)^\varepsilon = \varphi^\varepsilon * ( \tau_{ij}\lambda_j)$, we can rewrite the eqns. \eqref{eq:2_dens_w_vel_DR} in regularized form,
\begin{equation}
   \partial_t w_i^\varepsilon + \partial_j (u_j w_i)^\varepsilon - \frac{(w_i \theta)^\varepsilon}{2} = -\partial_i (\chi p)^\varepsilon + \partial_j (\chi \tau_{ij})^\varepsilon + (p \lambda_i)^\varepsilon - (\tau_{ij} \lambda_j)^\varepsilon\,,
   \label{eq:4_convol_NSE}
\end{equation}
where superscript $\varepsilon$ denotes convolution with $\varphi^\varepsilon$. This step regularizes the fields while preserving the structure of the equations, enabling a systematic analysis of scale interactions below and above $\varepsilon$. To derive the kinetic energy balance equation that combines $w_i$ and $w_i^{\varepsilon}$, we multiply the original equation \eqref{eq:2_dens_w_vel_DR} by $w_i^\varepsilon$ and the regularized equation \eqref{eq:4_convol_NSE} by $w_i$, then sum both to
\begin{equation}
\begin{aligned}
\partial_t (w_i w_i^\varepsilon) &+ \partial_j ( u_j w_i w_i^\varepsilon + w_j^\varepsilon \chi p - w_i^\varepsilon \chi \tau_{ij}) = \\ 
 & - w_i \partial_j ( u_j w_i)^\varepsilon + u_j w_i \partial_j w_i^\varepsilon + w_i \frac{\left( w_i \theta \right)^\varepsilon}{2} + \frac{w_i^\varepsilon w_i \theta}{2} \\ 
 & - w_i \partial_i \left( \chi p \right)^\varepsilon + \chi p \partial_i w_i^\varepsilon + w_i \left( p \lambda_i \right)^\varepsilon + w_i^\varepsilon p \lambda_i \\ 
 & + w_i \partial_j \left( \chi \tau_{ij} \right)^\varepsilon - \chi \tau_{ij} \partial_j w_i^\varepsilon - w_i \left( \tau_{ij} \lambda_j \right)^\varepsilon - w_i^\varepsilon \tau_{ij} \lambda_j \,. 
\end{aligned}
\label{eq:4_Regularized_NS}
\end{equation}
Furthermore, we can summarize some terms to,
\begin{equation}
    \begin{aligned}
   \int_{\mathbb{R}^3} \partial_j \varphi^\varepsilon (\bm{\xi})  \delta w_i \delta w_i \delta u_j\;d^3 \bm{\xi} =& -\partial_j \left( u_j w_i w_i \right)^\varepsilon + u_j \partial_j \left( w_i w_i \right)^\varepsilon - w_i w_i \partial_j u_j^\varepsilon \\
   &+ 2 w_i \partial_j \left( u_j w_i \right)^\varepsilon - 2 w_i u_j \partial_j w_i^\varepsilon\,,\\ 
   \int_{\mathbb{R}^3}\partial_i \varphi^\varepsilon (\bm{\xi})  \delta w_i \delta \chi \delta p\; d^3 \bm{\xi}= & -\partial_i \left( p w_i \chi \right)^\varepsilon + \chi \partial_i \left( p w_i \right)^\varepsilon + p \partial_i \left( w_i \chi \right)^\varepsilon - w_i \chi \partial_i p^\varepsilon - w_i p \partial_i \chi^\varepsilon\\
   & - p \chi \partial_i w_i^\varepsilon + w_i \partial_i \left( p \chi \right)^\varepsilon \,,\\ 
   \int _{\mathbb{R}^3}\partial_j \varphi^\varepsilon (\bm{\xi}) \delta w_i \delta \chi \delta \tau_{ij}\;d^3 \bm{\xi}=& -\partial_j \left( \tau_{ij} w_i \chi \right)^\varepsilon + \chi \partial_j \left( \tau_{ij} w_i \right)^\varepsilon + \tau_{ij} \partial_j \left( w_i \chi \right)^\varepsilon - w_i \chi \partial_j \tau_{ij}^\varepsilon - w_i \tau_{ij} \partial_j \chi^\varepsilon\\
   & - \tau_{ij} \chi \partial_j w_i^\varepsilon + w_i \partial_j \left( \tau_{ij} \chi \right)^\varepsilon\,, 
\end{aligned}
\label{eq:5_integral_relations}
\end{equation}
with $\delta{\bm f}({\bm \xi})={\bm f}({\bm x}+{\bm \xi})-{\bm f}({\bm x})$ the function increments over a distance vector ${\bm \xi}$, and $\bm f=(\bm w, \bm u, \chi, p, \tau_{ij})$. We will frequently use the shorthand notation of $\delta {\bm f}$ to ease the longer mathematical expressions. Substituting the last two terms from \eqref{eq:5_integral_relations} into \eqref{eq:4_Regularized_NS} leads to
\begin{equation}
    \begin{aligned}
   \partial_t (w_i w_i^\varepsilon) &+ \partial_j \left[ u_j w_i w_i^\varepsilon + w_j^\varepsilon \chi p - w_i^\varepsilon \chi \tau_{ij} + \frac{\left( u_j w_i w_i \right)^\varepsilon}{2} + \left( p w_j \chi \right)^\varepsilon - \left( \tau_{ij} w_i \chi \right)^\varepsilon \right] = \\ 
   &-\frac{1}{2}\int \partial_j \varphi^\varepsilon (\bm{\xi})  \delta w_i \delta w_i \delta u_j\,d^3 \bm{\xi} - \int \partial_i \varphi^\varepsilon (\bm{\xi})  \delta w_i \delta \chi \delta p\, d^3 \bm{\xi} + \int \partial_j \varphi^\varepsilon (\bm{\xi}) \delta w_i \delta \chi \delta \tau_{ij}\,d^3 \bm{\xi} \\ 
 & + \frac{1}{2}\left[ u_j \partial_j \left( w_i w_i \right)^\varepsilon - w_i w_i \partial_j u_j^\varepsilon + w_i \left( w_i \theta \right)^\varepsilon + w_i^\varepsilon w_i \theta\right] \\ 
 & + \chi \partial_i \left( p w_i \right)^\varepsilon + p \partial_i \left( w_i \chi \right)^\varepsilon - w_i \chi \partial_i p^\varepsilon - w_i p \partial_i \chi^\varepsilon + w_i \left( p \lambda_i \right)^\varepsilon + w_i^\varepsilon p \lambda_i \\ 
 & - \chi \partial_j \left( \tau_{ij} w_i \right)^\varepsilon - \tau_{ij} \partial_j \left( w_i \chi \right)^\varepsilon + w_i \chi \partial_j \tau_{ij}^\varepsilon + w_i \tau_{ij} \partial_j \chi^\varepsilon - w_i \left( \tau_{ij} \lambda_j \right)^\varepsilon - w_i^\varepsilon \tau_{ij} \lambda_j \,.
\end{aligned}
\label{eq:6_regularized_with_int}
\end{equation}
With further transformations, the expression \eqref{eq:6_regularized_with_int} results in
\begin{equation}
    \begin{aligned}
   \partial_t (w_i w_i^\varepsilon) &+ \partial_j [ u_j w_i w_i^\varepsilon + \chi (w_j^\varepsilon p + w_j p^\varepsilon) - \chi (w_i^\varepsilon \tau_{ij} + w_i \tau_{ij}^\varepsilon)+ \\
   &+ \frac{1}{2}\left[\left( u_j w_i w_i \right)^\varepsilon - u_j \left( w_i w_i \right)^\varepsilon\right] + \left( p w_j \chi \right)^\varepsilon - \chi \left( p w_j \right)^\varepsilon - \left( \tau_{ij} w_i \chi \right)^\varepsilon + \chi \left( \tau_{ij} w_i \right)^\varepsilon ] \\ 
  = &-\frac{1}{2}\int \partial_j \varphi^\varepsilon (\bm{\xi})  \delta w_i \delta w_i \delta u_j\,d^3 \bm{\xi} - \int \partial_i \varphi^\varepsilon (\bm{\xi})  \delta w_i \delta \chi \delta p\,d^3 \bm{\xi} + \int \partial_j \varphi^\varepsilon (\bm{\xi}) \delta w_i \delta \chi \delta \tau_{ij}\,d^3 \bm{\xi} \\ 
 & + \frac{1}{2}\left[-\left( w_i w_i \right)^\varepsilon \theta + w_i \left( w_i \theta \right)^\varepsilon + w_i \left( w_i^\varepsilon \theta - w_i \partial_j u_j^\varepsilon \right) \right] \\ 
 & - \left( p w_i \right)^\varepsilon \lambda_i + w_i \left( p \lambda_i \right)^\varepsilon + p \left( w_i^\varepsilon \lambda_i - w_i \partial_i \chi^\varepsilon \right) + p \partial_i \left( w_i \chi \right)^\varepsilon + p^\varepsilon \partial_i w_i \chi \\ 
 & + \left( \tau_{ij} w_i \right)^\varepsilon \lambda_j - w_i \left( \tau_{ij} \lambda_j \right)^\varepsilon - \tau_{ij} \left( w_i^\varepsilon \lambda_j - w_i \partial_j \chi^\varepsilon \right) - \tau_{ij} \partial_j \left( w_i \chi \right)^\varepsilon - \tau_{ij}^\varepsilon \partial_j w_i \chi. 
\end{aligned}
\label{eq:7_regularized_with_int_simpl}
\end{equation}
In the limit of $\varepsilon\to 0$, many terms in \eqref{eq:7_regularized_with_int_simpl} compensate each other. The term $(w_i w_i^\varepsilon)/2$ converges, for example, to the kinetic energy $E$. In this limit, we obtain the local kinetic energy balance equation:
\begin{equation}
\partial_t E + \partial_j \left( u_j E \right) = u_i \partial_j \tau_{ij} - u_j \partial_j p - \lim_{\varepsilon \to 0} \left( D_{wwu} + D_{w\chi p} - D_{w\chi \tau} \right).
\label{eq:8_local_kin_en_balance}
\end{equation}
The dissipation terms $D$ on the right-hand side are defined as follows,
\begin{subequations}
\begin{align}
   D_{wwu}(\varepsilon, {\bm x}) &= \frac{1}{4} \int \partial_j \varphi^\varepsilon (\bm{\xi})  \delta w_i \delta w_i \delta u_j\;d^3 \bm{\xi}\label{eq:9_diss_term_a}\\ 
   D_{w\chi p}(\varepsilon, {\bm x}) &= \frac{1}{2} \int \partial_i \varphi^\varepsilon (\bm{\xi})  \delta w_i \delta \chi \delta p\; d^3 \bm{\xi},\label{eq:9_diss_term_b}\\ 
   D_{w\chi \tau}(\varepsilon, {\bm x}) &= \frac{1}{2} \int \partial_j \varphi^\varepsilon (\bm{\xi}) \delta w_i \delta \chi \delta \tau_{ij}\; d^3 \bm{\xi}\label{eq:9_diss_term_c}.
\end{align}
\label{eq:9_diss_term}
\end{subequations}
These terms represent further energy dissipation terms in the compressible DR. The first two terms will be related to anomalous dissipation. Term \eqref{eq:9_diss_term_a} is due to velocity-velocity coupling ($D_{wwu}$), term \eqref{eq:9_diss_term_b} due to pressure-density-velocity interactions ($D_{w\chi p}$), and term \eqref{eq:9_diss_term_c} due to viscous stress-density-velocity coupling. They do not vanish for sufficient roughness of the velocity, pressure, and density fields in the inertial subrange for $\varepsilon\to0$, i.e. for scales $\eta_k\ll\xi$. In dimensionless form with Reynolds number $Re$ and Mach number $M$, the local kinetic energy balance thus becomes 
\begin{equation}
\partial_t E + \partial_j (u_j E) = - \frac{1}{\gamma M^2} u_j \partial_j p +\frac{1}{Re} u_i \partial_j \tau_{ij} - \lim_{\varepsilon \to 0} \left[ D_{wwu} + \frac{1}{\gamma M^2} D_{w\chi p} - \frac{1}{Re} D_{w\chi \tau} \right]\,.
\end{equation}
The dimensionless parameters are given by
\begin{equation}
M=\frac{U_{\rm rms}}{a_s} \quad\mbox{and}\quad Re=\frac{\rho_0 U_{\rm rms}L_0}{\mu}\,.
\end{equation}
Here, $L_0$ represents an outer scale of turbulence, e.g., the integral length scale \cite{Davidson2004}, and $U_{\rm rms}=\sqrt{\langle u_i^2\rangle_{V,t}}$ is the root mean square velocity, which measures the strength of velocity fluctuations. $a_s$ is the speed of sound and $\rho_0$ is a reference density, e.g., the mean density $\rho_0=\langle \rho\rangle_{V,t}$. For constant density $\rho({\bm x},t)=\rho_0$,  the terms $D_{w\chi p}$ and $D_{w\chi \tau}$ vanish since the density increment is zero $\delta\chi=0$. Moreover, the viscous work term simplifies to $u_i \partial_j \tau_{ij} =\partial^2_j(u_i^2/2) - \left( \partial_j u_i \partial_j u_i \right)$. Thus, the local kinetic energy balance \eqref{eq:8_local_kin_en_balance} simplifies to the incompressible kinetic energy balance with a single dissipation term \cite{Duchon2000,Zinchenko2024}. Thus, we recover the incompressible DR case,
\begin{equation}
\label{kin_en_eq}
    \partial _t E+ \partial_j(u_j E) = -u_j\partial_j p + \frac{1}{Re} [\Delta E - (\partial_j u_i)^2]-\underset{\varepsilon \to 0}{\mathop{\lim }}D_{wwu}\,,
\end{equation}
where $D_{wwu}$ corresponds to the anomalous dissipation derived by Duchon and Robert \cite{Duchon2000}, termed $D({\bm u})$ in the introduction. This confirms the generality of the framework.

\section{Framework by Aluie \cite{Aluie2013}}
\label{Sec::3}

In the study of turbulent flows, different frameworks have been developed to capture the physical mechanisms responsible for energy transfer and dissipation. The DR approach emphasizes energy dissipation arising from a lack of smoothness in weak solutions of the Euler and Navier–Stokes equations, while the framework developed by Aluie \cite{Aluie2013}, which we are going to outline in the following, is developed for the analysis of kinetic energy transfer across scales in compressible turbulence by using coarse-graining or Favre filtering \cite{Favre1969}. Starting point are the continuity and compressible Navier-Stokes equations in standard form,
\begin{equation}
\begin{aligned}
  \partial_t \rho + \partial_j (\rho u_j) &= 0\,,\\
  \partial_t (\rho u_i) + \partial_j (\rho u_j u_i) &= -\partial_i p + \partial_j \tau_{ij} + \rho f_i\,, 
\label{eq:NSE_comp}
\end{aligned}
\end{equation}
where the viscous stress tensor is given by \eqref{eq:1_dens_w_vel}. All flow variables are decomposed into large-scale (filtered) and small-scale components by applying a spatial filter at scale $\varepsilon$. Commuting the filter operation \eqref{eq:3_test_func} with space derivatives in the momentum and continuity equations yields
\begin{equation}
\begin{aligned}
  \partial_t \overline{\rho} + \partial_j \overline{\rho u_j} &= 0\,,\\
  \partial_t \overline{\rho u_i} + \partial_j \overline{\rho u_j u_i} &= -\partial_i \overline{p} + \partial_j \overline{\tau}_{ij} + \overline{\rho f_i}\,,
\end{aligned}
\label{eq:12_Aluie_1st_filtering}
\end{equation}
where overbar denotes spatial filtering $\overline{\bm a}=\varphi^\varepsilon * \bm a=\bm a^\varepsilon$. Even though this procedure is again a kind of regularization, we introduce a separate notation to distinguish this routine from DR. The scale decomposition employed in the large-scale momentum balance \eqref{eq:12_Aluie_1st_filtering} is equivalent to traditional filtering in large-eddy simulations (see, for example, \cite{Pope2000}). For the velocity and Reynolds stress components, which are denoted by $a$, a Favre filter, i.e., density-weighted averaging, leads to 
\begin{equation}
    \widetilde{a}(\varepsilon,{\bm x})=\frac{\overline{\rho a}}{\overline \rho}.
\label{eq:13_Aluie_dens_weighted}
\end{equation}
Since viscous effects are dominant at small scales, a spatial filtering averages all small-scale variations out. Thus, Favre filtering causes viscous effects at the large scales to be negligible. Now, using density-weighted parameters \eqref{eq:13_Aluie_dens_weighted}, we can rewrite momentum balance in the following form
\begin{equation}
  \overline{\rho}\partial_t  \widetilde{u}_i + \overline{\rho} \widetilde{u}_j\partial_j  \widetilde{u}_i = -\partial_j [\overline{\rho}\,\widetilde{\sigma} (u_i, u_j)] - \partial_i \overline{p} + \partial_j \overline{\tau}_{ij} + \overline{\rho} \widetilde{f}_i 
  \label{eq:14_Aluie_mom_balance}
\end{equation}
with 
\begin{equation}
    \overline{\rho}\,\widetilde{\sigma} (u_i, u_j) = \overline{\rho} \left( \widetilde{u_j u_i} - \widetilde{u}_j \widetilde{u}_i \right).
\end{equation}
Multiplying \eqref{eq:14_Aluie_mom_balance} by density-weighted velocity $\widetilde{u}_i$ and substituting the continuity equation, which is presented in the following form,
\[
\overline{\rho} \partial_t \widetilde{u}_i^2 = \partial_t (\overline{\rho}\,\widetilde{u}_i^2) + \partial_j (\overline{\rho}\,\widetilde{u}_i^2 \widetilde{u}_j) - 2\overline{\rho}\, \widetilde{u}_i \widetilde{u}_j \partial_j \widetilde{u}_i\,,
\]
into the momentum equation, one gets the following local kinetic energy balance \cite{Aluie2013} 
\begin{equation}
\begin{aligned}
   \partial_t\frac{\overline{\rho}\,\widetilde{u}_i^2}{2} + \partial_j J_j &= -\Pi - \Lambda + \overline{p} \partial_i \overline{u}_i - \overline{\tau}_{ij} \partial_j \widetilde{u}_i + \overline{\rho} \widetilde{f}_i \widetilde{u}_i, \\ 
  J_j(\varepsilon,{\bm x}) &= \left( \overline{\rho}\,\widetilde{u}_j \frac{\widetilde{u}_i^2}{2} + \bar{p}\,\bar{u}_j + \overline{\rho}\,\widetilde{u}_i\,\widetilde{\sigma}(u_i, u_j) - \widetilde{u}_i \overline{\tau}_{ij} \right), \\ 
  \Pi(\varepsilon,{\bm x}) &= -\overline{\rho}\,\widetilde{\sigma}(u_i, u_j) \partial_j \widetilde{u}_i, \\ 
  \Lambda(\varepsilon,{\bm x}) &= \frac{1}{\overline{\rho}}\,\overline{\sigma}(\rho, u_j) \partial_j \overline{p}, \\ 
  \overline{\sigma}(\rho, u_j)(\varepsilon,{\bm x}) &= \left( \overline{\rho u_j} - \bar{\rho} \bar{u}_j \right), \\ 
  \overline{\rho}\,\widetilde{\sigma}(u_i, u_j)(\varepsilon,{\bm x}) &= \overline{\rho} \left( \widetilde{u_j u_i} - \widetilde{u}_j \widetilde{u}_i \right).
\end{aligned}
\label{eq:20_Aluie_eq}
\end{equation}
Here $\overline{\rho}$, $\overline{p}$, and $\tilde{\bm{u}}$ are the Favre-averaged density, pressure, and velocity fields coarse-grained at a scale $\varepsilon$, respectively. Furthermore, $J_j$ is the spatial flux of the large-scale kinetic energy, $\overline{\tau}_{ij} \partial_j \widetilde{u}_i$ the viscous dissipation rate, and $\overline{\rho} \widetilde{f}_i \widetilde{u}_i$ the energy injection rate at large scales due to external (volume) forcing ${\bm f}$. 

Furthermore, the term $\overline{p}\partial _i\overline{u}_i$ is the pressure-dilatation at the resolved scale. It represents energy exchange between resolved kinetic energy and internal energy. This term does not contribute to energy transfer between scales. It is given by the product of the large-scale pressure and the large-scale dilatation. In incompressible Navier-Stokes flow, $\bar p \partial_i \bar u_i=0$. However, in the compressible case, it can become significant.

The term $\overline {\tau}_{ij}\partial \tilde u_i$ is the viscous dissipation, the rate at which molecular viscosity directly dissipates kinetic energy of scales $>\varepsilon$ into heat. For high-Reynolds-number flows, $\overline {\tau}_{ij}\partial \tilde u_i$ is essentially zero for all but the very finest scales; it can be shown rigorously \cite{Aluie2013} that viscous dissipation vanishes in the limit of $\varepsilon$ well above the dissipation scale of the flow.

The two remaining terms $\Pi(\varepsilon,{\bm x})$ and $\Lambda(\varepsilon,{\bm x})$ on the right-hand side of eq.\eqref{eq:20_Aluie_eq} are the subgrid-scale kinetic energy fluxes – they act as sinks of kinetic energy in the large-scale kinetic energy balance and thus as a source in the complementary small-scale balance. The term $\Pi(\varepsilon,{\bm x})$ is the {\em deformation work}, analogous to the turbulent energy flux in incompressible flow. It arises from large-scale velocity gradients that work against the subgrid stress. In this term, $\partial_j \tilde{u}_i$ is the resolved strain-rate tensor (on scales larger than $\varepsilon$) and $\tilde \sigma(u_i, u_j)$ is the turbulent stress. Physically, $\Pi(\varepsilon,{\bm x})$ represents the part of kinetic energy transferred downscale by the action of the large-scale strain and turbulent stress. A positive $\Pi$ means that large scales lose kinetic energy, which is passed down to smaller ones. This term arises from nonlinear interactions.

The quantity $\Lambda(\varepsilon,{\bm x})$ is termed {\em baropycnal work}, a scale-to-scale kinetic energy transfer mechanism that exists only in compressible flows where the density is scale-dependent. Here $\bar \sigma(\rho, u_j) =\overline{\rho u_j} - \bar{\rho} \bar{u}_j $ is the subgrid mass flux. The baropycnal work term $\Lambda(\varepsilon,{\bm x})$ represents the work done by the large-scale pressure-gradient force on the small-scale mass flux. In other words, it transfers energy across scales due to variations in pressure and density, thus a "baropycnal” work. Physically, $\Lambda$ captures effects similar to baroclinicity in the energy budget, as discussed in \cite{Aluie2019}. Clearly, this term vanishes in the incompressible limit and becomes significant when density gradients interact with pressure gradients. Together, $\Pi$ and $\Lambda$ constitute the total energy flux at scale $\varepsilon$. They are the only two processes through which kinetic energy can be transferred directly from small to large scales across $\varepsilon$ (and vice versa) in a variable-density flow. Both terms are sign-definite.

In summary, the coarse-grained kinetic energy balance in AL separates the effects of inter-scale energy transfer from the rest. The deformation work $\Pi$ and baropycnal work $\Lambda$ are the two channels of kinetic energy cascade across scales, analogous to the turbulent energy flux in incompressible turbulence, with $\Lambda$ being a uniquely compressible contribution. Meanwhile, the pressure-dilatation term appears as a distinct term in the large-scale energy equation, representing the conversion between kinetic and internal energy at the resolved scales. We are going to compare DR and AL in the next section. Thus, we provide the dimensionless form of the local kinetic energy balance at the end of this section, which is given by
\begin{equation}
\begin{aligned}
   \partial_t\frac{\overline{\rho}\,\widetilde{u}_i^2}{2} + \partial_j J_j &= -\Pi - \frac{1}{\gamma M^2}\Lambda + \frac{1}{\gamma M^2}\overline{p} \partial_i \overline{u}_i - \frac{1}{Re}\overline{\tau}_{ij} \partial_j \widetilde{u}_i + \frac{1}{F^2}\overline{\rho} \widetilde{f}_i \widetilde{u}_i, \\ 
  J_j(\varepsilon,{\bm x}) &= \left( \overline{\rho}\,\widetilde{u}_j \frac{\widetilde{u}_i^2}{2} + \frac{1}{\gamma M^2}\bar{p}\,\bar{u}_j + \overline{\rho}\,\widetilde{u}_i\,\widetilde{\sigma}(u_i, u_j) - \frac{1}{Re}\widetilde{u}_i \overline{\tau}_{ij} \right).
\end{aligned}
\label{eq:20_Aluie_eq_dimensionless}
\end{equation}
Here, Mach number $M$ and Reynolds number $Re$ are defined in the same way as in DR. An additional dimensionless parameter, $F=U_{\rm rms}/\sqrt{|{\bm f}| L_0}$ measures the ratio of inertial to external forcing in turbulent flow.

\section{Direct comparison of both frameworks}\label{Sec::4}

By comparing the local kinetic energy balance derived from frameworks DR \eqref{eq:8_local_kin_en_balance} and AL \eqref{eq:20_Aluie_eq}, we can identify analogous terms with respect to anomalous dissipation. They are compactly summarized in table \ref{tab:work_comparison} to initiate the comparison. Both approaches rely on spatial filtering with the same test function, a compact-support convolution kernel $\varphi^\varepsilon$, and capture the same physical mechanisms of kinetic energy transfer in compressible turbulence. However, they differ slightly in how these mechanisms are expressed and derived. 
\begin{table}[h]
\renewcommand{\arraystretch}{1.5}
\centering
\begin{tabular}{lcc}
\hline\hline
 & $\quad$ Duchon-Robert framework (DR) $\quad$ & $\quad$ Aluie framework (AL) $\quad$ \\
\hline
Deformation work & $-D_{wwu}(\varepsilon,{\bm x})$ & $-\Pi(\varepsilon,{\bm x})-\partial_j[\overline{\rho}\,\widetilde{u}_i \widetilde{\sigma}(u_i, u_j)](\varepsilon,{\bm x})$ \\
Baropycnal work & $-D_{w\chi p}(\varepsilon,{\bm x})$ & $-\Lambda(\varepsilon,{\bm x})$ \\
Viscous work & $D_{w\chi \tau}(\varepsilon,{\bm x}) + u_i \partial_j \tau_{ij}({\bm x})$ & $\widetilde{u}_i \partial_j \overline{\tau}_{ij}(\varepsilon,{\bm x})$ \\
\hline\hline
\end{tabular}
\caption{Comparison of local kinetic energy balance terms related to anomalous and viscous dissipation  between the frameworks DR (left column) and AL (right column).}
\label{tab:work_comparison}
\end{table}
In the following, we show in detail the correspondence between the main terms of both frameworks and discuss their physical meaning.
\begin{enumerate}
  \item {\em Deformation Work (Velocity coupling).} The anomalous term $D_{wwu}$ in DR stands for the inertial energy flux from large scales to small scales in the coarse-graining approach. Aluie represented the inertial transfer by the deformation work $\Pi$ with an additional contribution arising from the density variations of the energy flux. In the coarse-grained framework, the energy flux is expressed through a divergence term $\partial_j [\overline{\rho} \widetilde{u}_i \widetilde{\sigma}(u_i, u_j)]$, which can be rewritten as $\widetilde{u}_i \widetilde{\sigma}(u_i, u_j) \partial_j \overline{\rho} + \overline{\rho} \widetilde{u}_i \partial_j \widetilde{\sigma}(u_i, u_j)$. It represents the flow of filtered energy across the scale $\varepsilon$. This flux can be rewritten in a form that expresses the effect of density gradients more explicity,
\begin{equation*}
    \partial_j \widetilde{\sigma}(u_i, u_j) = \left( 2 \widetilde{u}_i \widetilde{u}_j - \widetilde{u_j u_i} \right) \frac{\partial_j \overline{\rho}}{\overline{\rho}} + \frac{\partial_j \overline{\rho u_j u_i}}{\overline{\rho}} - \widetilde{u}_i \frac{\partial_j \overline{\rho u_j}}{\overline{\rho}} - \widetilde{u}_j \frac{\partial_j \overline{\rho u_i}}{\overline{\rho}}\,.
\end{equation*} 
It can be seen that the inertial energy transfer in a variable-density flow has two contributions; one is $\Pi$, associated with the strain-rate work, analogous to the incompressible case. The other part is solely related to compressibility effects and in the absence of density variations, the divergence of the flux of stresses vanishes. In DR, both of these effects are covered by the single term $D_{wwu}$. It represents the nonlinear transfer of kinetic energy that remains finite in the limit $\varepsilon \to 0$. Physically, $D_{wwu}$ and $\Pi$ describe the same process, namely kinetic energy that is transferred from large eddies to smaller ones. When density is constant, this term reduces to the usual subgrid scale stress. When density varies, the extra flux contribution account for (additional) energy fluxes that appear due to compressibility. In summary, the quantity $D_{wwu}$ corresponds to the sum of the deformation work $\Pi$ and the energy flux related to density gradient in the coarse-grained framework. These represent inter-scale kinetic energy transfer.

\item {\em Baropycnal Work (Pressure–Density Coupling).} The term $D_{w\chi p}$ in DR corresponds directly to the baropycnal work term, which is denoted by $\Lambda$ in AL. Both, $D_{w\chi p}$ and $\Lambda$, arise from fluctuations in pressure and density. In physical terms, this is the work done by pressure forces in a variable-density flow that transfers kinetic energy across scales. In AL, $\Lambda$ appears as a separate contribution to the energy balance, representing how large-scale pressure gradients acting on small-scale density fluctuations. In DR, $D_{w\chi p}$ is defined by an integral of the product of three increments, $\delta w,\delta \chi,\delta p$, see eq. \eqref{eq:9_diss_term_b}. The term explicitly couples variations in the density field ($\chi = 1/\sqrt{\rho}$) with pressure and velocity and should not vanish in the limit of $\varepsilon\to 0$ if the fields are sufficiently rough. For example, this can be the case near shock fronts or strong compressions, indicating anomalous dissipation due to compressible effects. Importantly, the baropycnal term has no counterpart in incompressible turbulence, showing how compressibility introduces a new way cascading energy.

\item {\em Viscous work (Stress-Density Coupling).} Viscous effects appear differently in both  frameworks, but they are equivalent. In DR, the viscous dissipation in the kinetic energy equation comes from two terms, the regular viscous work $u_i \partial_j \tau_{ij}$ on the right side of eq. \eqref{eq:8_local_kin_en_balance} and the additional term $D_{w\chi \tau}$, which involves density and velocity fluctuations with the viscous stress, see eq. \eqref{eq:9_diss_term_c}. The sum $u_i\partial_j \tau_{ij} + D_{w\chi \tau}$ represents the total rate at which kinetic energy is converted into internal energy by viscosity. In AL, one finds a comparable term $\widetilde{u}_i\partial_j \overline{\tau}_{ij}$ in the filtered energy balance, which is the viscous dissipation at scale $\varepsilon$, the work of the viscous stress $\overline{\tau}_{ij}$ against the filtered velocity $\widetilde{u}_i$. Thus, $\widetilde{u}_i\partial_j \overline{\tau}_{ij}$ in AL corresponds to $u_i\partial_j \tau_{ij}$ plus the anomalous subgrid dissipation $D_{w\chi \tau}$. Both frameworks agree in the point that viscous forces are a sink of kinetic energy in the turbulent cascade. However, DR explicitly displays how much of this dissipation can be attributed to small-scale stress–velocity correlations with term $D_{w\chi \tau}$ when the solution is not smooth.
\end{enumerate}

Beyond identifying these corresponding terms, it is useful to show the structural differences between both frameworks. First, both frameworks employ an explicit filtering operation using a kernel $\varphi^\varepsilon$ to separate scales. In DR, the convolution smoothens the fields and one considers the limit of the filter scale of $\varepsilon\to 0$ to isolate singular contributions. In AL, the filter scale is kept finite to analyse energy fluxes at that scale. Only in the limit of very large Reynolds numbers $Re$ with an increasingly extended inertial range, one might consider $\varepsilon \to 0$.

Despite this difference, the filtering function can be chosen identically in both approaches, e.g., as a Gaussian function or any smooth compact-support mollifier normalized to 1 as done in \cite{Zinchenko2024} for the incompressible case. In fact, mathematically one can express filtered derivatives in the same way for both frameworks. For instance, an $n$th-order spatial derivative of a filtered field can be written as an integral over the field increments,
\begin{equation}
     \partial_j^n \overline{\bm{a}} = (-1)^n \int \partial_j^n \varphi^\varepsilon (\bm{\xi})   \bm{a}(\bm{r}+\bm{\xi}) d^3 \bm{\xi}=(-1)^n \int \partial_j^n \varphi^\varepsilon (\bm{\xi})   (\bm{a}(\bm{r}+\bm{\xi})-\bm{a}(\bm{r})) d^3 \bm{\xi}, \quad n\in \mathbb{N}.
\end{equation}
This identity is exactly the same as the relation used to derive both the anomalous terms in DR and AL; it shows that both frameworks are based on the same filtering formalism. Notably, for the DR, we use a density-weighted velocity $w_i=\sqrt{\rho}u_i$ to simplify the kinetic energy density, whereas AL uses Favre filtering for velocity, $\widetilde{u}_i = \overline{\rho u_i}/\overline{\rho}$. These are mathematically related choices to handle the variable density. Despite the different notation, the study of compressibility effects through filtering is thus consistent between the two methods.

The Duchon–Robert framework yields a local kinetic energy balance that is exact in the distributional sense. It states that locally there can be “energy leakage” through the anomalous dissipation terms $D_{wwu}$, $D_{w\chi p}$, and $D_{w\chi \tau}$ if the fields are not smooth. These terms are essentially zero for smooth solutions and become non-zero only when the solution develops sufficiently rough fields. Specifically, the terms $D_{wwu}$ and $D_{w\chi p}$ represent anomalous dissipation and $D_{w\chi \tau}$ is an additional contribution to the viscous dissipation arising from compressibility effects.

Both frameworks point to three key physical processes in compressible turbulence: (1) the "classical" anomalous dissipation term is captured by $D_{wwu}$ in DR and by $\Pi$ in AL, (2) the second dissipation term, baropycnal work, is governed by pressure-density fluctuations and captured by $D_{w\chi p}$ and $\Lambda$. (3) The last term, viscous dissipation, is captured by $D_{w\chi \tau}$ and the filtered viscous term $\widetilde{u}_i \partial_j \overline{\tau}_{ij}$, and represents kinetic energy losses to internal energy. Both approaches show a consistent physical picture. Large-scale kinetic energy is passed down to smaller scales by rough velocity, pressure, and density fields, and finally, viscosity at the smallest scales converts the kinetic energy into thermal energy. 

In conclusion, the Duchon–Robert and the Aluie frameworks for compressible turbulence are in close agreement on the nature of energy transfer and dissipation. They apply the same filtering methodology and equivalent physical terms can be identified in their balances: deformation work with compressible effects, baropycnal work, and viscous dissipation. The primary difference stems from their formulation. Despite this difference, for an incompressible flow case, both frameworks reduce to the classic incompressible energy cascade result. In the following, we are going to analyse these terms from both frameworks for a simple one-dimensional flow.  

\section{One-dimensional compressible flow examples and anomalous dissipation analysis}\label{Sec::5}

To calculate and compare the anomalous dissipation terms of both frameworks, we study two one-dimensional (1D) flow configurations. First, we look at a fully compressible 1D flow as in a simple gas dynamics process; secondly, at an adiabatic 1D flow to show the difference in the evolution of flow parameters and dissipation. We consider 1D fully compressible Navier–Stokes equations for an ideal gas with spatial coordinate $x$ and time $t$. The balances of mass, momentum, and energy together with the equation of state in physical dimensions are given by
\begin{equation}
\begin{aligned}
  \partial_t \rho + \partial_i (\rho u_i) &= 0, \\ 
  \rho \partial_t u_i+\rho u_j\partial_j u_i &= -\partial_i p + \mu \partial^2_k u_i + \frac{\mu}{3} \partial_i (\partial_k u_k), \\ 
  \rho C_p d_t T - \beta T d_t p &= \mu \Phi_{diss} + \lambda \partial^2_k T + q_v, \\ 
  p &= \rho RT. 
\end{aligned}
\label{eq:11_fully_compres_flow}
\end{equation}
with $d_t=\partial_t+u_j\partial_j$. Here, $T$ is the temperature field, $R$ is the ideal gas constant, and $\beta$ is the volumetric expansion coefficient. All material parameters, dynamic viscosity $\mu$, thermal conductivity $\lambda$, and specific heat at constant pressure, $C_p$, are assumed to be constant with respect to $x$ and $t$. Furthermore, $\Phi_{diss} = ( \partial_k u_l + \partial_l u_k - \frac{2}{3} (\partial_m u_m) \delta_{kl} ) \partial_k u_l$,  $q_v = 0$, and  $\beta \approx 1/T$. For the one-dimensional case, the equations simplify to
\begin{equation}
\begin{aligned}
  \partial_t \rho + \partial_x( \rho u) &= 0\,,\\
  \rho\partial_t u + \rho u \partial_x u &= -R \partial_x( \rho T) + \frac{4}{3}\mu \partial_{xx} u\,, \\ 
  \rho C_v ( \partial_t T + u \partial_x T) + \rho R T \partial_x u &= \mu \frac{4}{3} ( \partial_x u)^2 + \lambda \partial_{xx} T\,. 
\end{aligned}
\label{eq:12_1D_fully_compres_flow}
\end{equation}
The dimensionless form with Mach and Reynolds numbers is given by
\begin{equation}
\begin{aligned}
  \partial_t \rho + \partial_x ( \rho u) &= 0\,,\\
  \rho \partial_t u + \rho u \partial_x u &= -\frac{1}{\gamma M^2} \partial_x(\rho T) + \frac{4}{{3Re}} \partial_{xx} u\,, \\ 
  \partial_t T + u \partial_x T + \left( \gamma - 1 \right) T \partial_x u &= \frac{4M^2}{{3Re}} \gamma \left( \gamma - 1 \right) \frac{\left( \partial_x u \right)^2}{\rho} + \frac{\gamma}{Pr {Re}} \frac{\partial_{xx} T}{\rho}\,.
\end{aligned}
\label{eq:13_1D_dimensionless_compres_flow}
\end{equation}
Here, $\gamma=C_p/C_v$ is the adiabatic constant and $Pr=\mu C_p/\lambda$ is the Prandtl number.
We take a periodic velocity profile with constant density and temperature distributions as the initial conditions,
\begin{equation}
    \label{eq:14_simple_in_cond}
    \left\{ 
    \begin{aligned}
        u\left( x,0 \right)&=\sin \left( 2\pi x \right) \\ 
        \rho \left( x,0 \right)&=1 \\ 
        T\left( x,0 \right)&=1 \\ 
    \end{aligned} 
    \right.
\end{equation}
Any pressure or density changes at $t>0$ will be generated by the motion of the gas. In our numerical approach for solving these equations numerically, we employ a fourth-order Runge–Kutta method for time integration. Spatial derivatives are computed using an eighth-order compact finite-difference scheme, which ensures precise resolution of steep gradients and shock features. The domain is discretized by $10^4$ uniform mesh cells, which results in a grid spacing of $dx=10^{-4}$, while the time integration is performed over $3\times 10^5$ steps from $t=0$ to 1.5 time units, resulting in a time step of $dt=5 \cdot 10^{-6}$. Periodic boundary conditions are applied for computation at the ends of the one-dimensional interval. This high-order scheme is specifically chosen to accurately capture the steep gradients and complex features of compressible flow, enhancing both the stability and precision of the simulation. For more detailed information about the numerical scheme, see also Appendix \ref{Appendix_A}.

\begin{figure}[h]
\centering
    \includegraphics[width=0.8\linewidth]{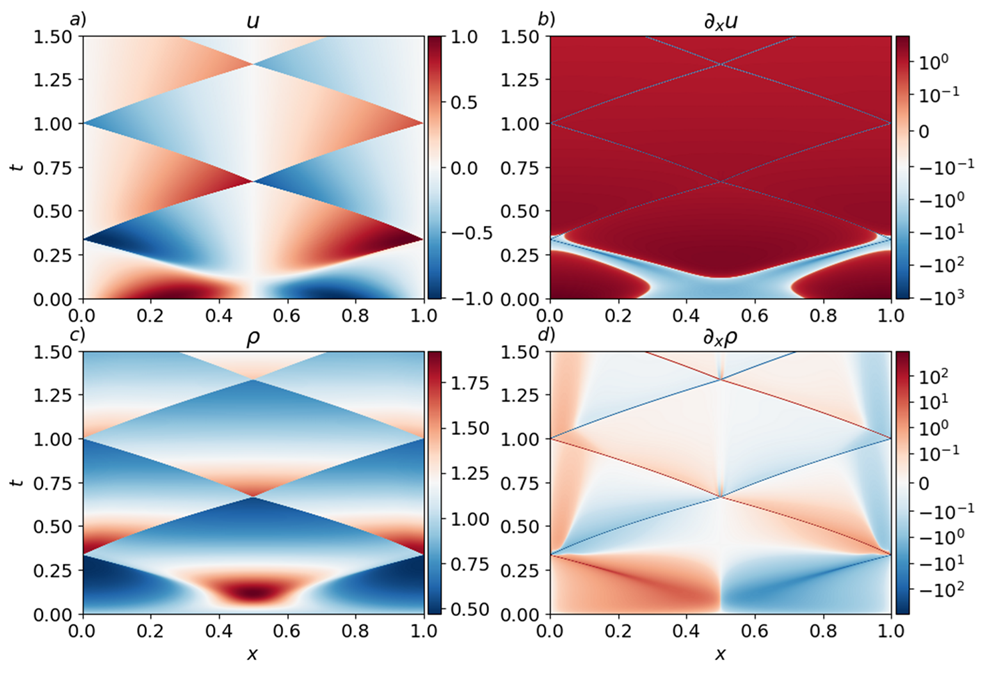}
    \caption{Evolution of velocity and density fields as well as their derivatives in 1D compressible flow. Each panel is a space–time diagram ($x$ on the horizontal axis, $t$ on the vertical axis) for a specific variable. (a) Velocity $u(x,t)$, (b) velocity derivative $\partial_x u(x,t)$, (c) density $\rho(x,t)$, and finally (d) density derivative $\partial_x \rho(x,t)$. The shocks meet at $t\approx 0.34$ at $x=0$ (and $x=1$ due to periodic boundary conditions) and at $t\approx 0.67$ at $x=0.5$ and so on. Here, $Re=5000$, $\gamma=1.4$, and $M=0.7$.}
    \label{Fig::1_simple_cond}
\end{figure}

Figure \ref{Fig::1_simple_cond} shows how the initial velocity distribution and the other fields evolve, in particular, how shock waves form, cross each other, and thus interact. 
The simple initial velocity profile consists of a single broad perturbation, which leads to the formation of two symmetric shock waves. As the simulation begins, the fluid in the perturbed region accelerates and compresses the gas ahead of it, resulting in the formation of shock waves. We observe in Figure \ref{Fig::1_simple_cond} that after a short time $t\approx0.34$, sharp discontinuities in $u$ and $\rho$ develop. These are indicated as thin red/blue lines in the $\partial_x u$ and $\partial_x \rho$ panels, indicating the presence of shock fronts. The two generated shock waves propagate in opposite directions. The formed shock waves pass by each other every time interval $\Delta t\approx 0.34$ due to periodic boundary conditions. At this moment, the two waves merge, which leads to the appearance of a new shock that carries the combined compression and strength of both initial waves, resulting in a velocity jump and almost negligible density and pressure gradients. After this short moment of merging, the two shock waves instantly split, returning to the previous amplitudes of velocity, density, and pressure and moving in opposite directions until the next time they merge. Over time, due to non-vanishing viscosity, the strength of the shock waves decreases until the shock waves completely disappear. 

\begin{figure}[h]
\centering
    \includegraphics[width=1\linewidth]{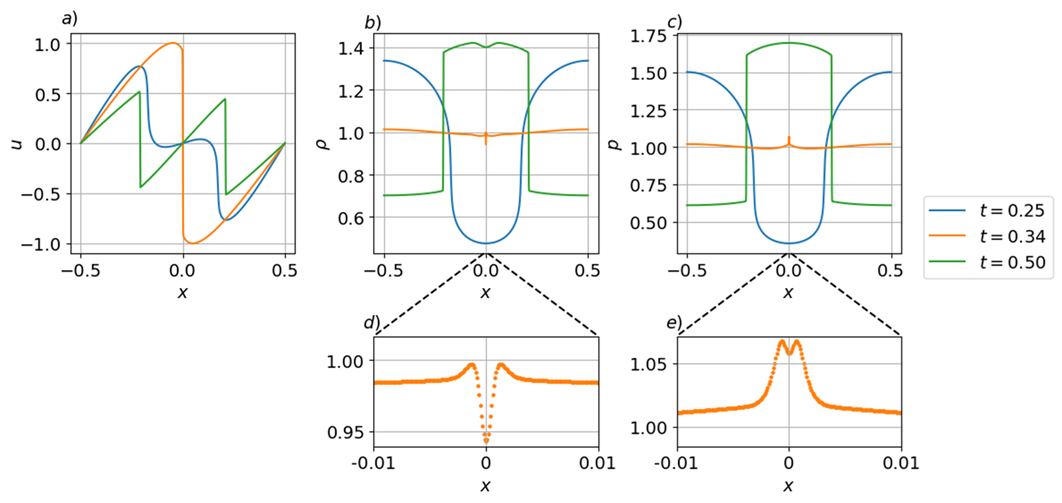}
    \caption{Spatial profiles of (a) velocity, (b) density, and (c) pressure at the time instant of shock wave crossing. The bottom row contains magnifications of the vicinity of the merging region to highlight the cusp-like profile for (e) density and (f) pressure. Parameters are the same as in Fig. \ref{Fig::1_simple_cond}.}
    \label{Fig::2D_distrib}
\end{figure}

A more detailed spatial profile of fields at the shock interaction instant is presented in Fig. \ref{Fig::2D_distrib}. While the velocity develops steep gradients, the density and pressure gradients are much weaker. However, a different type of non-smooth behavior can develop during a merging event of the shock fronts, namely, a cusp in the density and pressure fields. For example, a discontinuity $\delta\rho(x\approx0,\xi) \sim |\xi|^h$ is seen in the orange curve of density (and pressure). During the shock crossing event, the minimum value of Hölder exponent $h_{min} \approx 0.15$. It means that the baropycnal work, see eq.\eqref{eq:9_diss_term_b}, involves interactions between different types of singularities, sharp shocks with nearly discontinuous velocity gradients, and local regions where density and pressure have Hölder-type distributions.

\subsection{Fully compressible flow case}\label{Sec::6}
Using the 1D flow configuration described above, we now apply the kinetic energy balance \eqref{eq:8_local_kin_en_balance} and calculate the anomalous dissipation terms \eqref{eq:9_diss_term}. With the same dimensionless parameters as in equation \eqref{eq:13_1D_dimensionless_compres_flow}, the local kinetic energy balance equation for $E=\rho u^2/2$ in dimensionless form is given by
\begin{equation}
\partial_t E + \partial_x (u E) = u\left(\frac{4}{3Re} \partial_{xx} u -\frac{1}{\gamma M^2}  \partial_x p\right) - \underset{\varepsilon \to 0}{\lim} \left( D_{wwu} +\frac{1}{\gamma M^2} D_{w\chi p} -\frac{1}{Re} D_{w\chi \tau} \right),
\label{eq:16_dimensionless_kin_en}
\end{equation}
Individual dissipation contributions are now defined as follows 
\begin{equation}
\begin{aligned}
   D_{wwu}(\varepsilon,x) &= \frac{1}{4 \varepsilon^2} \int \psi_1 \delta w^2 \delta u \, d \xi, \\
   D_{w\chi p}(\varepsilon,x) &= \frac{1}{2 \varepsilon^{2}} \int \psi_1 \delta w \delta \chi \delta p \, d \xi, \\
   D_{w\chi \tau}(\varepsilon,x) &= \frac{1}{2 \varepsilon^{2}} \int \psi_1 \delta w \delta \chi \delta \tau_{11} \, d \xi,
\end{aligned}
\label{eq:18_3_diss_terms}
\end{equation}
with $w(x,t)=\sqrt{\rho}u$ and $\tau_{11}(x,t)=(4/3) \partial_x u$. Here, $\varepsilon$ again defines the filter or coarse-grain scale. The dimensionless test function $\psi_1$ is selected using theory of wavelets \cite{Mallat1999}. Generally, they are given by  
\begin{equation}
\psi_m(\xi) = (-1)^{m+1} \,\partial_x^m [\exp(-\xi^2 / 2)].
\label{eq:27_gausian_wavelet}
\end{equation}
To calculate the dissipation term in \eqref{eq:16_dimensionless_kin_en}, we have to use the 1st-order Gaussian wavelet which is given by
\begin{equation}
    \psi_1(\xi) = \partial_x [\exp\left(-\xi^2/2\right)] = -\xi \exp(-\xi^2/2)\,,
    \label{eq:1dwavelet}
\end{equation}
which provides a smooth filter to measure kinetic energy transfer across scale $\varepsilon$.

\begin{figure}[h]
\centering
\includegraphics[width=0.99\linewidth]{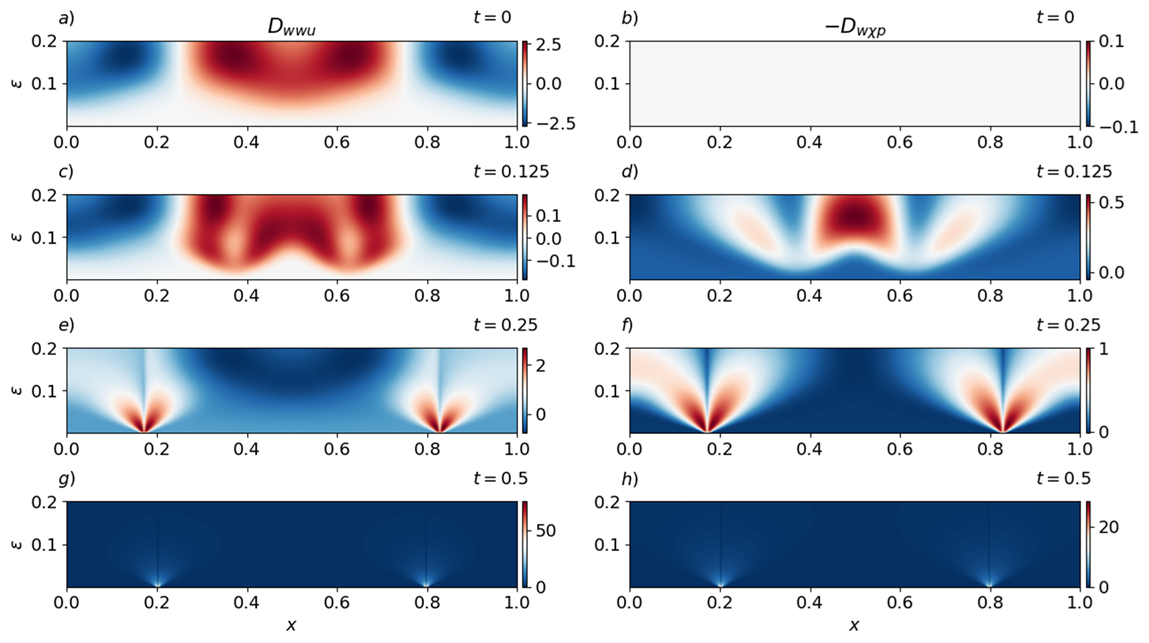}
    \caption{Temporal evolution of the anomalous dissipation terms $D_{wwu}(\varepsilon,x)$ (left column) and and $-D_{w\chi p}(\varepsilon,x)$ (right column) in DR. Snapshots are shown at different times to illustrate how these terms change and where they get enhanced in the flow. (a,b) $t=0$, (c,d) $t=0.125$, (e,f) $t=0.25$, and (g,h) $t=0.5$. The smallest value of $\varepsilon$, which is taken in the analysis in all panels, is $\varepsilon=10^{-4}$.}
    \label{Fig::Diss_in_time}
\end{figure}

\subsubsection{Analysis in the Duchon-Robert framework}

\begin{figure}[h]
\centering
\includegraphics[width=0.9\linewidth]{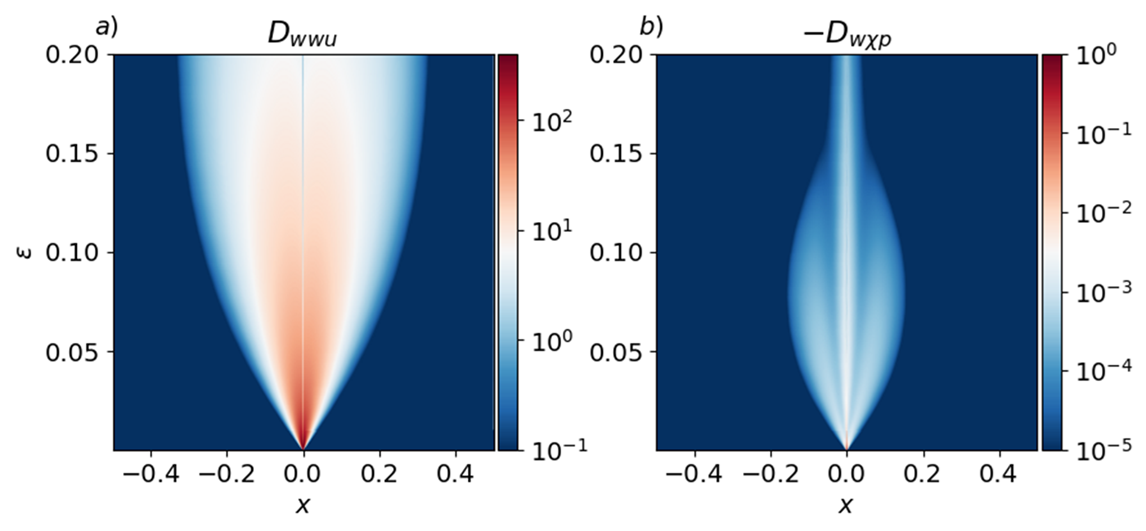}
    \caption{Dissipation near the shock interaction region. The time is now $t\approx 0.34$ when both shocks meet at $x=0$. Due to periodic boundary conditions, we continue the $x$ axis to the left with negative values. The interaction strongly influences the spatial and temporal distribution of both terms. Panel (a) $D_{wwu}(\varepsilon,x)$ and panel (b) $-D_{w\chi p}(\varepsilon,x)$.}
    \label{Fig::Diss_interaction}
\end{figure}

With the given parameters, the anomalous dissipation terms can be calculated for the initial condition \eqref{eq:14_simple_in_cond}. For the following, we fix the comparison study to a Reynolds number $Re=5000$, an adiabatic coefficient $\gamma=1.4$, and a Mach number $M=0.7$. The behavior of shocks is similar starting from time $t=0.5$, as it becomes periodic with decaying amplitudes. Therefore, for the dissipation analysis, we will focus only on the time interval $t\le 0.5$. In Fig. \ref{Fig::Diss_in_time}, the evolution of the dissipation terms  $D_{wwu}(\varepsilon,x)$ and $-D_{w\chi p}(\varepsilon,x)$ is shown at four different time instants. They have similar behavior when 2 shocks appear, but differ when these shocks interact. This interaction occurs periodically at times $t\approx 0.34$ and 0.67 at $x=0$ and 0.5, respectively, and continues alternating between both points. See again Fig. \ref{Fig::1_simple_cond}. At the interaction points, the velocity gradient remains large, while variations in density, pressure, and temperature are almost absent, as can be seen in Fig. \ref{Fig::Diss_interaction}, where we plot the dissipation terms. 
\begin{figure}[h]
\centering
\includegraphics[width=0.9\linewidth]{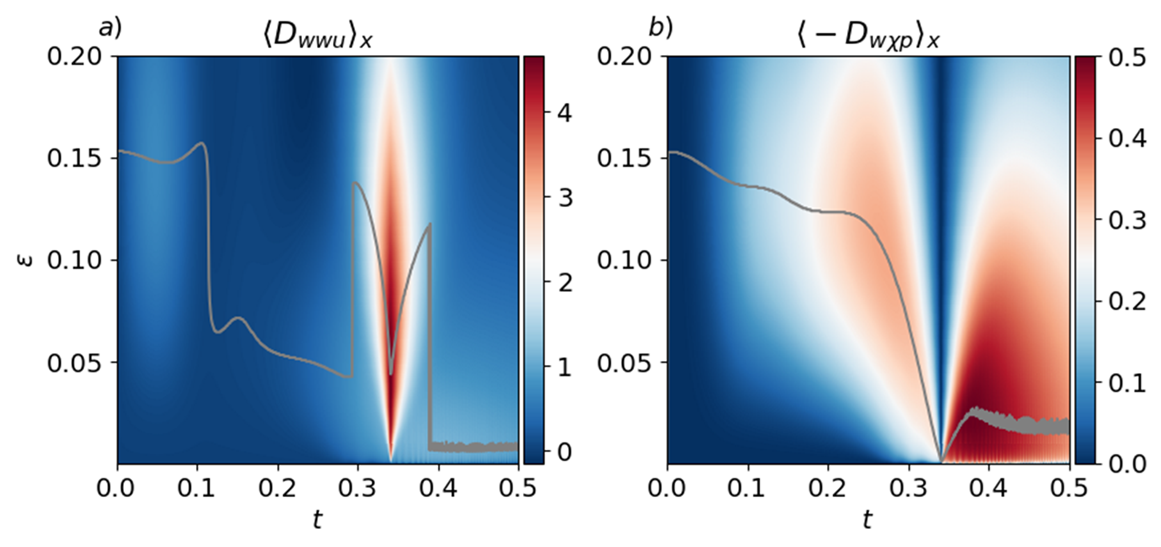}
    \caption{Spatially averaged anomalous dissipation terms versus time $t$ and filter scale $\varepsilon$. (a) $\langle D_{wwu}\rangle_x$, (b) and $\langle -D_{w\chi p}\rangle_x$. The gray line in both contour plots stands for $\max_{\varepsilon}\langle D\rangle_x$ at time $t$.}
    \label{Fig::Diss_mean}
\end{figure}
Both figures display symmetric contours, but at strongly different amplitudes. The dissipation in panel (b) of Fig. \ref{Fig::Diss_interaction} is almost absent, indicating weak coupling between velocity, density, and pressure. This corresponds to $t\approx 0.34$, when two shocks interact, and the density and pressure increments are nearly zero. In contrast, panel (a) of the same figure shows strong dissipation, highlighting the dominant nonlinear coupling of velocity.
\begin{figure}[h]
\centering
\includegraphics[width=0.99\linewidth]{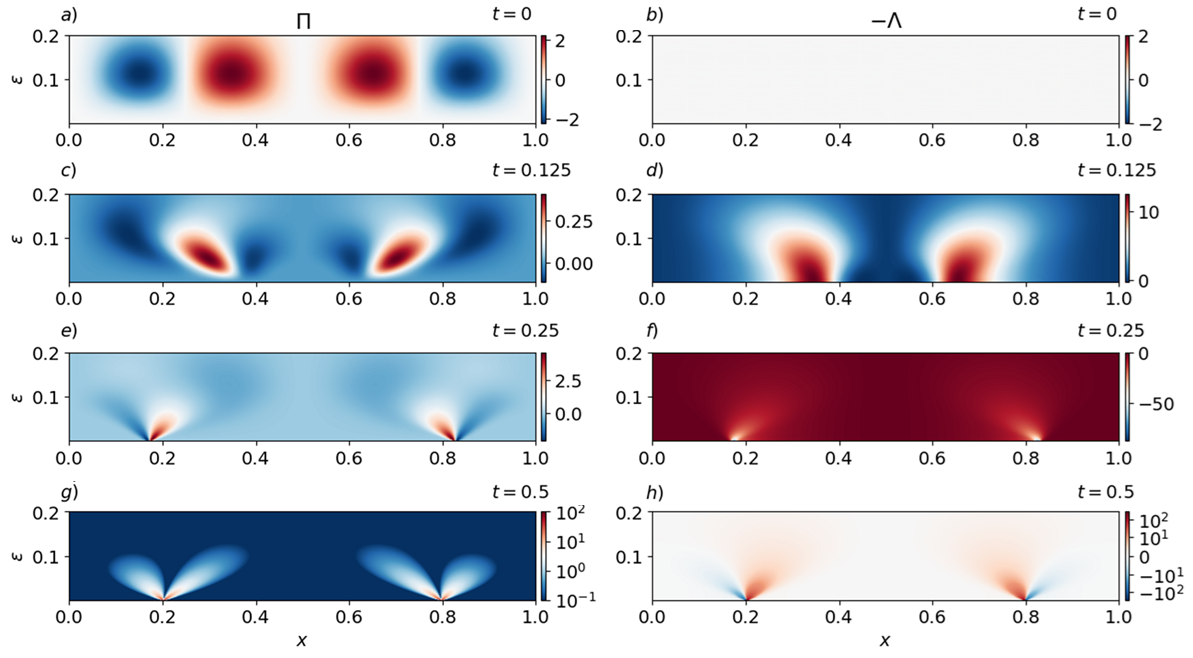}
    \caption{Temporal evolution of the deformation work $\Pi(\varepsilon,x)$ (left column) and the baropycnal work $\Lambda(\varepsilon,x)$ (right column) in AL. Snapshots are shown at different times to illustrate how these terms change and where they concentrate in the flow. (a,b) $t=0$, (c,d) $t=0.125$, (e,f) $t=0.25$, and (g,h) $t=0.5$. The figure has to be compared to Fig. \ref{Fig::Diss_in_time}. The smallest value of $\varepsilon$, which is taken in the analysis in all panels, is $\varepsilon=10^{-4}$.}
    \label{Fig::Aluie_diss}
\end{figure}
\begin{figure}[h]
\centering
\includegraphics[width=0.9\linewidth]{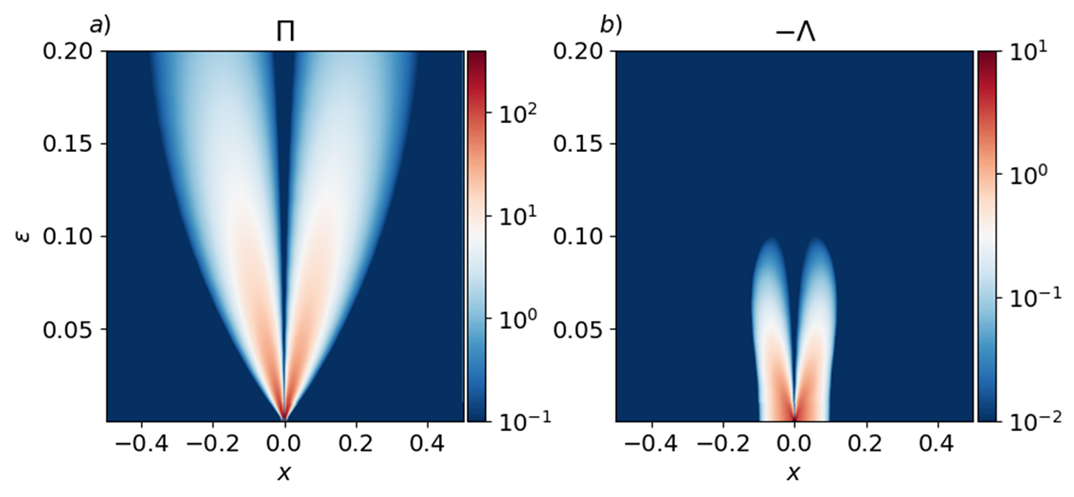}
    \caption{Deformation $\Pi (\varepsilon,x)$ in panel (a) and baropycnal work $\Lambda(\varepsilon,x)$ in panel (b) near the shock interaction region. The time is now $t\approx 0.34$ when both shocks meet at $x=0$. Due to periodic boundary conditions, we continue the $x$ axis to the left with negative values.}
    \label{Fig::Aluie_interaction}
\end{figure}

In order to simplify the analysis and subsequent comparisons with AL, it is useful to consider spatially averaged anomalous dissipation terms, $\langle D\rangle_x$. They are displayed in  Fig. \ref{Fig::Diss_mean} for the DR case.
\begin{figure}[h]
\centering
\includegraphics[width=0.9\linewidth]{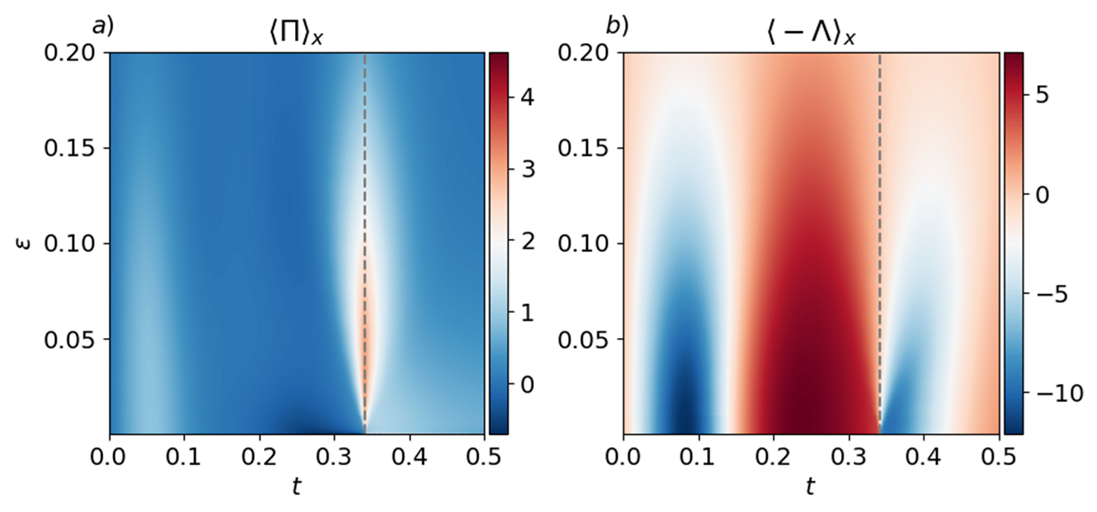}
    \caption{Spatially averaged work terms versus time $t$ and filter scale $\varepsilon$. (a) Deformation work $\langle\Pi\rangle_x$, (b) and baropycnal work $\langle -\Lambda\rangle_x$. The gray dashed line in both contour plots stands for the shock interaction instant of $t\approx 0.34$.}
    \label{Fig::Aluie_averaged}
\end{figure}
Both figures show the spatially averaged dissipation, with significantly higher magnitudes in panel (a) than in panel (b). The dissipation peaks around $t\approx 0.34$, where shock interactions occur. While the left panel displays a concentrated and intense peak at this time instant, the right panel shows an almost negligible value at this point and generally has a broader distribution. The gray lines in both panels stand for scale $\varepsilon$ at which the maximum dissipation occurs. For example, in panel (a) this is the case for $\varepsilon\approx 0.17$ at $t=0$.

\subsubsection{Comparison with the Aluie framework}

The newly obtained terms can be directly compared to AL. We therefore analyse the deformation work $\Pi(\varepsilon,x)$ and the baropycnal work $\Lambda(\varepsilon,x)$. In Fig. \ref{Fig::Aluie_diss}, their evolution over time is shown. Before the shocks interact, AL yields sharper peaks, making it easier to locate the regions of interest. However, once the shocks begin to interact, the results become qualitatively similar to those obtained from DR. In both approaches, $\Pi$ and $D_{wwu}$ remain mostly positive, indicating that energy is dissipated at a specific coarse-grain scale $\varepsilon$ due to strong velocity fluctuations. Meanwhile, baropycnal work $\Lambda$ generally reaches higher magnitudes than the deformation work $\Pi$, except in the shock interaction zone where they become comparable, as seen in Fig.~\ref{Fig::Aluie_interaction}. Here, the density and pressure fields are relatively smooth, causing baropycnal work to be much smaller than deformation work. 

Similarly to the DR case, it is also interesting to analyse the spatially averaged terms $\langle \Pi \rangle_x$ and $\langle \Lambda \rangle_x$. The result is displayed in Fig. \ref{Fig::Aluie_averaged}. Both panels exhibit different behaviors with respect to their DR counterparts. The left panel of the figure shows the deformation work, which has nearly the same behavior as the averaged dissipation term $\langle D_{wwu}\rangle_x$ in Fig. \ref{Fig::Diss_mean}(a), however with smaller magnitudes. In contrast, the contours in panel (b) differ significantly in comparison to $\langle -D_{w\chi p}\rangle_x$. The mean baropycnal work oscillates in time between positive and negative values, depending on the location of the shocks, and has in general bigger magnitudes than the deformation work. We observe notable changes near $t\approx 0.34$ that coincide with shock interactions. This time instant is indicated by the gray dashed line in both figures. In contrast to deformation work, baropycnal work $\Lambda$ changes sign exactly at this time instant.

\subsection{Adiabatic flow case}\label{Sec::5b}

\begin{figure}[h]
\centering
    \includegraphics[width=0.8\linewidth]{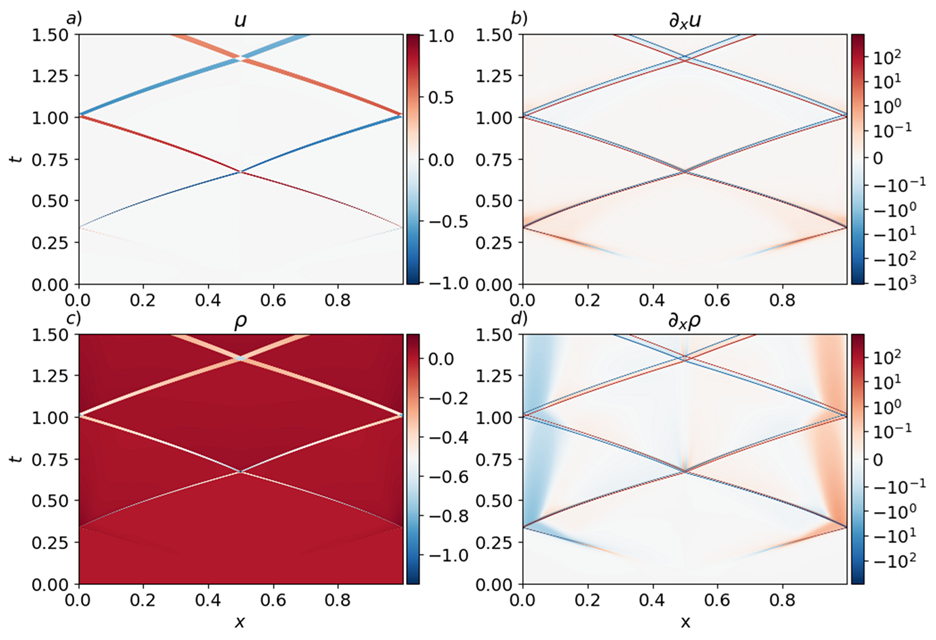}
    \caption{Difference between the evolution of the adiabatic and fully compressible flow cases at $Pr = 0.7$ for the simple initial condition \eqref{eq:14_simple_in_cond}.  Each panel is a space–time diagram ($x$ on the horizontal axis, $t$ on the vertical one) for a specific variable. (a) Velocity $u$, (b) velocity gradient $\partial_x u$, (c) density $\rho$, and (d) density gradient $\partial_x \rho$.}
    \label{diff_adiabat-fullu_compres}
\end{figure}

After exploring the fully compressible one-dimensional Navier–Stokes case, i.e., the case with a full internal energy balance, we now consider a simplified adiabatic flow case which is given by
\begin{equation}
    \begin{aligned}
  \partial_t \rho + \partial_x (\rho u) &= 0\,, \\ 
  \rho \partial_t u + \rho u \partial_x u &= -\partial_x p + \frac{4}{3} \mu \partial_{xx} u\,, \\ 
  \frac{p}{p_0} &= \left( \frac{\rho}{\rho_0} \right)^\gamma\,.
\end{aligned}
\end{equation}
The former energy equation is now closed by the adiabatic gas law. The adiabatic coefficient $\gamma=1.4$, as in the former example. The dimensionless form of these equations, with the pressure substituted from the adiabatic equation of state, can be written as
\begin{equation}
\begin{aligned}
  \partial_t \rho &= -\partial_x (\rho u)\,\\
  \partial_t u &= -u \partial_x u - \frac{\gamma}{M^2} \rho^{\gamma - 2} \partial_x \rho + \frac{4}{\mathrm{3\rho Re}}\partial_{xx} u\,.
\end{aligned}
\end{equation}
Figure \ref{diff_adiabat-fullu_compres} illustrates the difference between the adiabatic flow and the fully compressible case with respect to time. A value of zero means that both solutions predict the same value at that point, while nonzero colors indicate where they deviate. The results show that the differences between the adiabatic and fully compressible case are very small in most of the space-time domain; they are concentrated primarily around the shock waves. This also implies that in the smooth flow regions, heat flux does not play a significant role. 
It can be seen that the difference at the shock locations is characterized by the same amplitudes; furthermore, the overall evolution remains the same. The only difference is the velocity of shock wave propagation. At later times, a gradual widening of the colored bands is detected, which shows the distance between the shock waves in the adiabatic and fully compressible cases. As a result, the comparison in Figure \ref{diff_adiabat-fullu_compres} indicates that using the adiabatic flow configuration has only a minor effect on the overall flow dynamics. The primary impact of heat fluxes is to slightly smooth and weaken the shocks. We have varied the Prandtl number between $Pr=0.07$ and $Pr=7$ and could not detect a significant impact of the variation of this parameter.

\begin{figure}[h]
\centering
\includegraphics[width=0.8\linewidth]{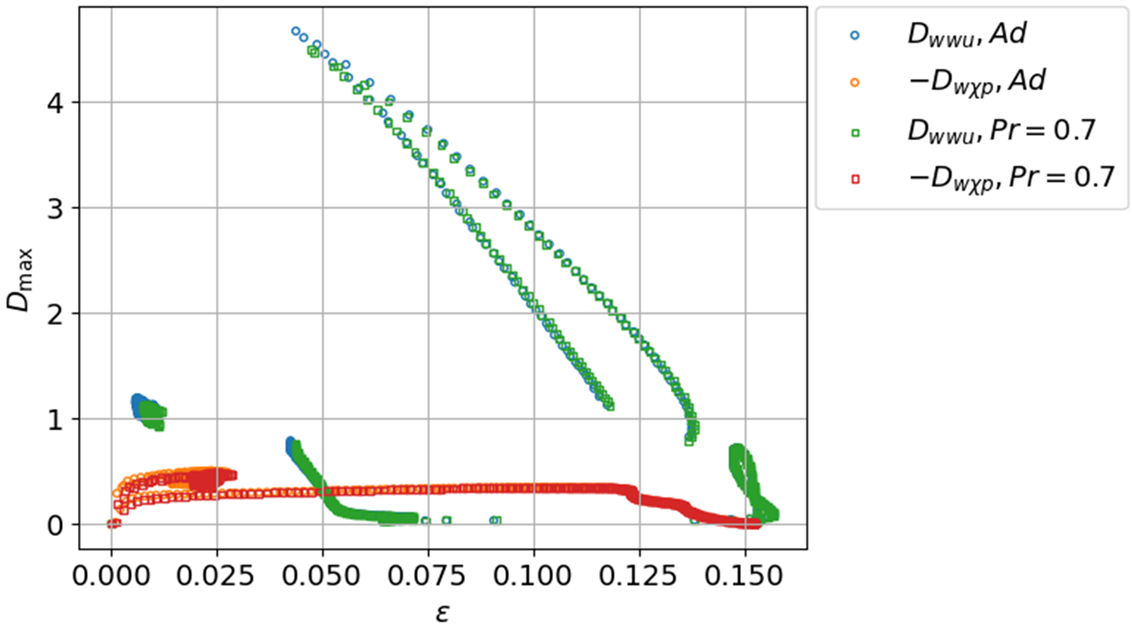}
    \caption{Maxima of the DR mean dissipation terms for the adiabatic (Ad) and the fully compressible ($Pr=0.7$) flow cases, denoted $D_{\rm max}$, versus coarse-grain scale $\varepsilon$. Points denote the maxima at different time instants in the course of the evolution for the different coarse-grain scales.}
    \label{Fig::Diss_max}
\end{figure}

It is also important to understand the influence of the adiabatic flow conditions on the anomalous dissipation terms in DR. For this reason, we take the mean dissipation values corresponding to the gray lines in both panels of Fig. \ref{Fig::Diss_mean}, and replot them together with the corresponding adiabatic ones in  Fig. \ref{Fig::Diss_max}. 
Blue and orange circles correspond to the case of the adiabatic flow, while green and red squares to the flow with $Pr=0.7$. For larger coarse-grain scales $\varepsilon>0.5$, there is almost no difference between the results (not shown). However, at smaller scales, the adiabatic flow shows slightly higher dissipation at the same scale than the real flow. In the adiabatic case, gradients cannot be smoothed by heat fluxes. As a result, the shocks are sharper and more localized, with steeper gradients in velocity and density. This results in greater roughness at small scales. The anomalous dissipation is sensitive to steep gradients and velocity increments $\sim \delta u^3$. Thus, we can conclude that the influence of adiabatic conditions in the one-dimensional case is nearly negligible, which means that it is possible to use these simplified conditions to analyse dissipation in the present one-dimensional flow example.

\section{Dissipation terms in an analytical shock configuration}\label{Sec::7}
Neither of the reviewed cases can fully capture what happens to the dissipation terms in the limit where the coarse-graining scale approaches zero, $\lim_{\varepsilon \to 0}[D_{wwu}(\varepsilon,x)+D_{w\chi p}(\varepsilon,x)+D_{w\chi\tau}(\varepsilon,x)]$. To better understand the limiting behavior, we consider an idealized shock-like velocity profile that approximates real flow conditions, as illustrated in Fig. \ref{Fig::1_simple_cond}. It is important to note that the following shock profile is not an exact solution to the compressible Navier–Stokes equations. This analytical profile captures a discontinuity in the velocity field, approximating two crossing shocks, as we have observed in the simulations. The velocity $u(x)$ is defined piecewise linear over a periodic domain $x\in[-0.5,0.5]$ as follows,
\begin{equation}
    u=\left\{ 
    \begin{aligned}
     &A(2x+1),\quad -0.5\le x\le-0.5+d;\\ 
     & 2Ax,\quad -0.5+d< x<0.5-d; \\ 
     & A(2x-1),\quad 0.5-d \le x\le 0.5. 
    \end{aligned} \right.
    \label{eq:piecewise}
\end{equation}
Here, $A$ represents the amplitude of the shock wave, determining its strength, and $d$ denotes the distance of the discontinuity from the periodic boundary. Now we can examine specific time instants, particularly when two shocks interact, such as at $t\approx0.34$ in the numerical examples, or for $d=0.5$ in this case. In the latter case, the middle segment of \eqref{eq:piecewise} vanishes and the profile simplifies to a single centered shock:
\begin{equation}
    u=\left\{ 
    \begin{aligned}
     &A(2x+1),\quad x<0;\\ 
     & A(2x-1),\quad x>0. 
    \end{aligned} \right.
\end{equation}
which jumps from $u(0^-)=A$ to $u(0^+)=-A$ at $x=0$. At this moment, pressure and density variations are negligible, as it was shown before, so the dissipation is dominated by the velocity increment term $D_{wwu} (\varepsilon,x)\approx \rho_c D_{uuu}(\varepsilon,x)$ as $\rho\approx\rho_c$. Thus, we discuss $D_{uuu}$ only for the following. The velocity increment $\delta u(x,\xi)=u(x+\xi)-u(x)$ for the piecewise linear shock profile becomes
\begin{equation}
    \delta u=-u+\left\{ 
    \begin{aligned}
     &A(2(x+\xi)+1),\quad \xi<-x;\\ 
     & A(2(x+\xi)-1),\quad \xi>-x. 
    \end{aligned} \right.
\end{equation}
The anomalous dissipation term can then be evaluated analytically as a convolution of the velocity increments with a Gaussian filter, see again eq. \eqref{eq:1dwavelet},
\begin{equation}
    D_{uuu}(\varepsilon,x)= -\frac{1}{4 \varepsilon^3} \int_{-\infty}^{+\infty} \xi \exp \left(\frac{\xi^2}{2\varepsilon^2}\right)(\delta u)^3 \, d \xi,
\end{equation}
At the exact location of the shock $x=0$, the result simplifies to
\begin{equation}
    D_{uuu}(\varepsilon,0)=\frac{A^3}{2}\left(\frac{1}{\varepsilon}-3\sqrt{2\pi}+24\varepsilon-12\sqrt{2\pi}\varepsilon^2\right)\,.
\end{equation}
For small $\varepsilon\ll 1$, the first term inside the brackets dominates. This implies that an infinitely steep shock, a discontinuous velocity jump, produces a dissipation term that infinitely grows as $\varepsilon \to 0$,
\begin{equation}
    \lim_{\varepsilon \to 0}D_{uuu}(\varepsilon,0) = \lim_{\varepsilon \to 0} \frac{1}{\varepsilon} = \infty.
\end{equation}
In other words, a shock with an infinitely sharp gradient shows an anomalous dissipation that tends to infinity for vanishing filter scale.

Away from the shock at $x \neq 0$ in the smooth regions, the dissipation results in the following respective functions to the left $x<0$ and to the right $x>0$ of the shock,
\begin{equation}
    \frac{D_{uuu}}{A^3}=\frac{2 + 12\varepsilon^2}{\varepsilon} \exp\left( -\frac{x^2}{2\varepsilon^2} \right) - 6\sqrt{2\pi} \varepsilon^2 + 6\frac{x^2 \pm x}{\varepsilon} \exp\left( -\frac{x^2}{2\varepsilon^2} \right) - 3\sqrt{2\pi} \left[1 \pm \mathrm{erf}\left( \frac{x}{\sqrt{2} \varepsilon} \right)\right]\,.
\end{equation}
One can verify that as $\varepsilon \to 0$, dissipation tends to zero for any fixed $x \neq 0$. Physically, the anomalous dissipation vanishes in regions where the flow is smooth (as it is expected). Meanwhile, at the shock position, it is extremely large in an infinitely thin region around $x=0$. If we compute the mean dissipation over one full periodic region, the divergent contribution at the shock has infinitely small spatial width, yielding a finite average. Integrating $D(x)$ over the whole domain and taking the limit $\varepsilon\to 0$ gives an average dissipation of
\begin{equation}
    \langle D_{uuu}\rangle_x=\frac{A^3}{2}\,,
\end{equation}
which is finite and independent of coordinate $x$ and filter scale $\varepsilon$. This simple example confirms that if locally dissipation at the ideal shock is singular, the mean energy dissipation rate remains finite.

\begin{figure}[h]
\centering
\includegraphics[width=0.6\linewidth]{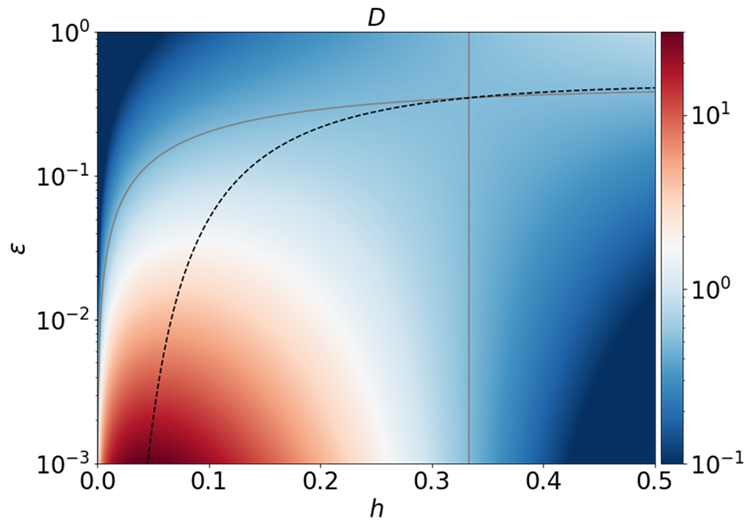}
    \caption{Dissipation for Hölder-type profile of the velocity field $u(x)$. The grey line shows the constant dissipation $D=1/2$ corresponding to the Hölder exponent $h=1/3$.  Black dashed line determines the maximum dissipation given by \eqref{Eq::47_eps_h_max}.}
    \label{Fig::11_Hoelder_distrib}
\end{figure}

The other way to show finite energy dissipation is to use a different type of singularity for the velocity profile. Therefore, we consider a profile of the form
\begin{equation}
    u(x)=A\left| x\right|^h\,,
\end{equation}
where $h$ is the Hölder exponent characterizing the steepness of the velocity singularity. While this profile does not correspond to the shock-wave configuration discussed earlier, it serves a different purpose. First, in compressible flow, density and pressure fields show Hölder-type distribution near shocks. Thus, this profile offers a simplified way to show the influence of such singularities on kinetic energy dissipation. Second, Hölder exponent is the threshold for the anomalous dissipation. It is known that finite anomalous dissipation occurs only when $h\le1/3$. This makes the Hölder-type velocity profile a natural test case for analyzing anomalous dissipation. At the point of singularity $x=0$, its increment becomes
\begin{equation}
    \delta u(x=0,\xi)=A\left| \xi\right|^h.
\end{equation}
As this form is symmetric, we cannot use the 1st-order Gaussian wavelet, as this results to a trivial zero value. We switch to the 2nd-order Gaussian wavelet, see eq. \eqref{eq:27_gausian_wavelet} for $m=2$. It is known as the "Mexican Hat" wavelet and given by
    \begin{equation}
    \bm{\psi}_2(\xi) =(1-\xi^2) \exp(-\xi^2/2)\,.
\end{equation}
Substituting the velocity increment and test function into the dissipation integral results to
\begin{equation}
    D_{uuu}(\varepsilon,0)= \frac{A^3}{2 \varepsilon^2} \int_{0}^{\infty} \left( 1-\frac{\xi^2}{\varepsilon^2}\right)\xi^{3h} \exp \left(-\frac{\xi^2}{2\varepsilon^2}\right) \, d \xi\,,
\end{equation}
which leads to 
\begin{equation}
    D(\varepsilon,h)=\frac{3}{2}A^3\frac{h}{\left( \sqrt2 \varepsilon\right)^{1-3h}}\Gamma\left(\frac{1+3h}{2}\right)\,,
\end{equation}
with the Gamma function $\Gamma(y)$ and $D_{uuu}(\varepsilon,x=0;h)=D(\varepsilon,h)$. The resulting function $D(\varepsilon,h)$ is shown as a heat map in Fig. \ref{Fig::11_Hoelder_distrib}
The integral is finite in the limit $\varepsilon\to0$ only when Hölder exponent $h=1/3$, leading to $D=A^3/2$; a Hölder exponent $h=1/3$ of the velocity leads to a finite anomalous dissipation locally. For $h<1/3$, we again receive local singularity. The smaller $h$ ($h>0$) the stronger the divergence. The roughness parameter $h$ for the strongest singularity can be determined through the maximum dissipation $d_h D(\varepsilon,h)=0$. This results to
\begin{equation}
    \varepsilon=\frac{1}{\sqrt 2}\exp\left[ -\frac{1}{3h_{\max}}-\frac{1}{2}\Gamma_p\left( \frac{1+3h_{\max}}{2}\right)\right],
\label{Eq::47_eps_h_max}
\end{equation}
where $\Gamma_p$(y) is a Polygamma function. For the smoother velocity case of $h>1/3$, the dissipation vanishes in the limit $\varepsilon\to 0$. If the velocity field is smoother than $|x|^{1/3}$ at $x=0$, then
\begin{equation}
    \lim_{\varepsilon \to 0}D(\varepsilon,h)=0\,,
\end{equation}
which implies that no anomalous energy dissipation occurs at the point of singularity. These findings are consistent with Onsager's theoretical criterion for energy conservation in ideal flows \cite{Onsager1949}. Onsager conjectured that a Hölder exponent of $h=1/3$ is the critical threshold separating dissipative and non-dissipative anomalous behavior in turbulent flows.

\begin{figure}[h]
\centering
\includegraphics[width=0.6\linewidth]{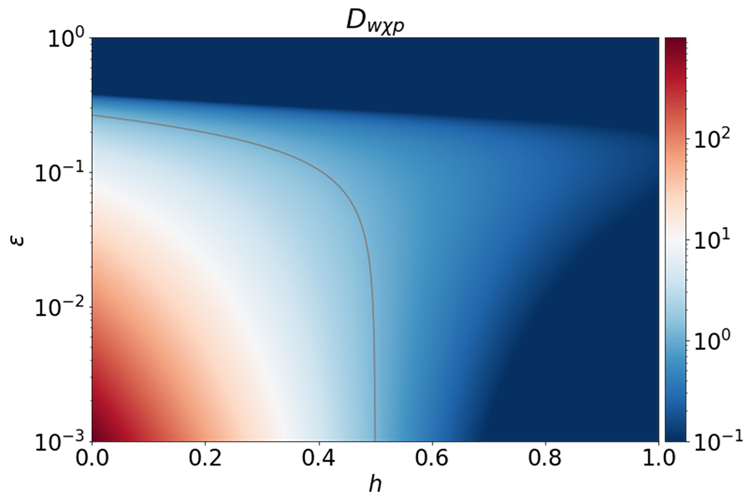}
    \caption{Baropycnal dissipation term with the shock-type distribution of the velocity field and Hölder-type distribution of the pressure and density field. The grey line shows the constant dissipation $D_{w\chi p}=\sqrt{\pi /2}$ corresponding to the Hölder exponent $h=1/2$.}
    \label{Fig::13_Hoelder_distrib_baropyc}
\end{figure}

In the first part of this section, we found that during shock merging, coupling of different types of singularities can happen. This influences baropycnal work which we want to discuss finally. We assume the following fields at this moment near $x\approx0$,
\begin{align}
  w &=\sqrt \rho u =
  \begin{cases}
    A(2x + 1), & x < 0 \\
    A(2x - 1), & x > 0
  \end{cases} \\
  \chi&=\frac{1}{\sqrt{\rho}} = a_{\chi} + b_{\chi} |x|^{h_1} \\
  p &= a_p + b_p |x|^{h_2}
\end{align}
Their increments at point $x=0$ are given by
\begin{align}
  \delta w &=
  \begin{cases}
    A(2\xi + 1), & \xi < 0 \\
    A(2\xi - 1), & \xi > 0
  \end{cases} \\
  \delta\chi&= b_{\chi} |\xi|^{h_1} \\
  \delta p &= b_p |\xi|^{h_2}
\end{align}
For this discussion, we again use the first-order test function, such that results to
\begin{equation}
    D_{w\chi p} (\varepsilon,0,h_1,h_2)= -\frac{b_\chi b_p}{2 \varepsilon^3} \int_{-\infty}^{+\infty}  \xi \exp \left(-\frac{\xi^2}{2\varepsilon^2}\right) \left|\xi\right|^{h_1+h_2}\delta w \, d \xi.
\end{equation}
This leads to
\begin{equation}
    D_{w\chi p}(\varepsilon,0,h_1,h_2)=Ab_{\chi}b_p\frac{1}{\left(\sqrt 2 \varepsilon \right)^{1-h_1-h_2}}\left[ \sqrt 2\Gamma\left(\frac{2+h_1+h_2}{2}\right)-4\varepsilon\Gamma\left(\frac{3+h_1+h_2}{2}\right)\right]\,.
\end{equation}
In case of the same roughness of the pressure $p$ and field $\chi$, $h_1=h_2=h$, the dissipation term remains finite for fields with Hölder exponent $h=1/2$, which is shown in a heat map in  Fig. \ref{Fig::13_Hoelder_distrib_baropyc}.

\section{Summary and Outlook}\label{Sec::8}
The purpose of this study was to identify mechanisms of anomalous dissipation for fully compressible turbulent flows. To this end, we extended the Duchon-Robert framework (DR) \cite{Duchon2000}, which is originally formulated for incompressible flows, to the case of a fully compressible flow. We then considered one-dimensional fully compressible flow examples, where the dissipation mechanism is essentially determined by the generation of shock waves and their crossings due to periodic boundary conditions.

The anomalous dissipation contributions, represented as different terms $D(\varepsilon,x)$, were determined within DR and compared with the framework that was suggested by Aluie (AL) in ref. \cite{Aluie2013}. We could relate dissipation terms of DR and AL to each other. The kinetic energy dissipation appears in the form of three additional integral terms in the local kinetic energy balance. While two terms, $D_{wwu}$ and $D_{w\chi p}$ can be directly identified as anomalous dissipation terms, the third one, $D_{w\chi\tau}$ contains a viscous stress contribution and thus does not represent an anomalous dissipation term. The limit of the coarse-grain or filter scale of $\varepsilon\to 0$ has to be taken to determine its magnitude. The dependency of $D(\varepsilon,x)$ on the finite filter scale $\varepsilon$ and on the position $x$ away from a shock position was investigated in two examples numerically using a high-order compact finite difference scheme. To this end, we considered two one-dimensional examples, a first case with an internal energy balance (termed the fully compressible case) and a second case for which the internal energy balance is closed by an adiabatic law for the state variables, $p\sim \rho^{\gamma}$. The analysis showed that the results of the two models exhibit qualitatively (as well as almost quantitatively) similar behavior for the anomalous dissipation terms. DR and AL terms for anomalous dissipation could be related to each other, similarities and differences, which arise from both frameworks, were discussed. 

Moreover, we approximated the shock profiles in a simple analytical model for the discontinuity to describe the behavior of the anomalous dissipation terms within the inertial subrange. The model profile was guided by the numerical simulation results. For the simple piecewise profile, our discussion showed that the anomalous dissipation increases as the filtering scale $\varepsilon\to 0$. For the more singular shock structures, dissipation grows at the finer scales, while the mean kinetic energy dissipation rate remains finite. Considering a Hölder continuous profile, it was shown that the dissipation remains finite even locally for $h=1/3$.

Possible extensions of the present study include (1) the enlargement of the consideration to two- and three-dimensional compressible flows, where shock surfaces and their interactions can result in higher orders of anomalous dissipation; and (2) the extension of the approach to fully compressible three-dimensional homogeneous isotropic turbulence, where additional dynamical mechanisms, such as the stretching of compressible vortices, generation due to baroclinic vorticity production, and shock-vortex interactions contribute to the energy cascade as well as dissipation. These studies are underway and will be reported elsewhere.

\acknowledgments
The work of GZ was supported by the Priority Programme DFG-SPP 2410 ``Hyperbolic Balance Laws in Fluid Mechanics: Complexity, Scales, Randomness (CoScaRa)`` funded by the Deutsche Forschungsgemeinschaft (DFG). We acknowledge discussions with Roshan Samuel and L\'{a}szl\'{o} Sz\'{e}kelyhidi.

\appendix
\section{Details on the numerical scheme}
\label{Appendix_A}
For time integration, we use the classic explicit fourth-order Runge–Kutta (RK4) method. This scheme evaluates the right-hand side four times $k_1,k_2,k_3,k_4$ to achieve overall fourth-order accuracy in time,
\begin{equation}
    \begin{matrix}
\begin{aligned}
  & u_{i}^{n+1}=u_{i}^{n}+\frac{\Delta t}{6}\left( k_{1u}+2k_{2u}+2k_{3u}+k_{4u} \right), \\ 
 & k_{1u}=f_u\left( t^n,u^n,\rho^n \right), \\ 
 & k_{2u}=f_u\left( t^n+\frac{\Delta t}{2},u^n+\frac{k_{1u}}{2},\rho^n+\frac{k_{1\rho}}{2} \right), \\ 
 & k_{3u}=f_u\left( t^n+\frac{\Delta t}{2},u^n+\frac{k_{2u}}{2},\rho^n+\frac{k_{2\rho}}{2} \right), \\ 
 & k_{4u}=f_u\left( t^n+\Delta t,u^n+k_{3u},\rho^n+k_{3\rho} \right),
\end{aligned}
&
\begin{aligned}
  & \rho_{i}^{n+1}=\rho_{i}^{n}+\frac{\Delta t}{6}\left( k_{1\rho}+2k_{2\rho}+2k_{3\rho}+k_{4\rho} \right), \\ 
 & k_{1\rho}=f_\rho\left( t^n,u^n,\rho^n \right), \\ 
 & k_{2\rho}=f_\rho\left( t^n+\frac{\Delta t}{2},u^n+\frac{k_{1u}}{2},\rho^n+\frac{k_{1\rho}}{2} \right), \\ 
 & k_{3\rho}=f_\rho\left( t^n+\frac{\Delta t}{2},u^n+\frac{k_{2u}}{2},\rho^n+\frac{k_{2\rho}}{2} \right), \\ 
 & k_{4\rho}=f_\rho\left( t^n+\Delta t,u^n+k_{3u},\rho^n+k_{3\rho} \right),
\end{aligned}
\end{matrix}
\end{equation}
Detailed descriptions of RK methods can be found in textbooks on numerical differential equations \cite{Butcher2016}.

For spatial derivatives, we apply a high-order compact finite-difference scheme. Compact schemes couple neighboring points to compute derivatives. In our case, an 8th-order compact scheme is used \cite{LELE1992}, which means a tridiagonal system has to be solved to find the derivative values at each point:
\begin{equation}
    \begin{aligned}
   \beta_{c}C_{i+2} + \alpha_{c}C_{i+1} + C_{i} + \alpha_{c}C_{i-1} &+ \beta_{c}C_{i-2} = c_{c}\frac{u_{i+3} - u_{i-3}}{6\Delta} + b_{c}\frac{u_{i+2} - u_{i-2}}{4\Delta} + a_{c}\frac{u_{i+1} - u_{i-1}}{2\Delta}, \\ 
 \beta_{v}V_{i+2} + \alpha_{v}V_{i+1} + V_{i} + \alpha_{v}V_{i-1} &+ \beta_{v}V_{i-2} = c_{v}\frac{u_{i+3} - 2u_{i} + u_{i-3}}{9\Delta^2} + b_{v}\frac{u_{i+2} - 2u_{i} + u_{i-2}}{4\Delta^2} +\\ &+a_{v}\frac{u_{i+1} - 2u_{i} + u_{i-1}}{\Delta^2}.
\end{aligned}
\end{equation}
Here $C$ and $V$ refer to the first- and second-order derivatives, respectively. Constants in the scheme are defined as,
\begin{equation*}
    \begin{aligned}
  & a_{c} = \frac{25}{16}, \quad b_{c} = \frac{1}{5}, \quad c_{c} = -\frac{1}{80}, \quad \alpha_{c} = \frac{3}{8}, \quad \beta_{c} = 0, \\
  & a_{v} = \frac{147}{152}, \quad b_{v} = \frac{51}{95}, \quad c_{v} = -\frac{23}{760}, \quad \alpha_{v} = \frac{9}{38}, \quad \beta_{v} = 0.
\end{aligned}
\end{equation*}

\bibliographystyle{plain}
\bibliography{references}

\begin{thebibliography}{10}

\bibitem{Aluie2013}
H.~Aluie.
\newblock Scale decomposition in compressible turbulence.
\newblock {\em Physica D}, 247:54--65, 2013.

\bibitem{Buaria2020}
D.~Buaria, A.~Pumir, and E.~Bodenschatz.
\newblock {Self-attenuation of extreme events in Navier-Stokes turbulence}.
\newblock {\em Nat. Commun.}, 11:5852, 2020.

\bibitem{Burgers1948}
J.~M. Burgers.
\newblock A mathematical model illustrating the theory of turbulence.
\newblock {\em Adv. Appl. Mech.}, 1:171--199, 1948.

\bibitem{Butcher2016}
C.~Butcher.
\newblock {\em Numerical Methods for Ordinary Differential Equations}.
\newblock John Wiley \& Sons, Ltd., Chichester, England, 2016.

\bibitem{Davidson2004}
P.~A. Davidson.
\newblock {\em Turbulence: An Introduction for Scientists and Engineers}.
\newblock Oxford University Press, Oxford, 2004.

\bibitem{Laszlo2014}
C.~De~Lellis and L.~Székelyhidi.
\newblock {Dissipative Euler flows and {Onsager's }conjecture}.
\newblock {\em J. Eur. Math. Soc.}, 16:1467--1505, 2014.

\bibitem{Dubrulle2019}
B.~Dubrulle.
\newblock {Beyond Kolmogorov cascades}.
\newblock {\em J. Fluid Mech.}, 867:P1, 2019.

\bibitem{Duchon2000}
J.~Duchon and R.~Robert.
\newblock {Inertial energy dissipation for weak solutions of incompressible
  Euler and Navier-Stokes equations}.
\newblock {\em Nonlinearity}, 13:249--255, 2000.

\bibitem{Eyink2018}
G.~L. Eyink and T.~D. Drivas.
\newblock Cascades and dissipative anomalies in compressible fluid turbulence.
\newblock {\em Phys. Rev. X}, 8:011022, 2018.

\bibitem{Favre1969}
A.~J. Favre.
\newblock {\em Statistical equations of turbulent gases, in Problems of
  Hydrodynamics and Continuum Mechanics}.
\newblock Society for Industrial and Applied Mathematics, Philadelphia, PA,
  1969.

\bibitem{Frisch1995}
U.~Frisch.
\newblock {\em Turbulence: The Legacy of A. N. Kolmogorov}.
\newblock Cambridge University Press, Cambridge, UK, 1995.

\bibitem{Hamlington2008}
P.~E. Hamlington, J.~Schumacher, and W.~J.~A. Dahm.
\newblock {Direct assessment of vorticity alignment with local and nonlocal
  strain rates in turbulent flows}.
\newblock {\em Phys. Fluids}, 20:{111703}, 2008.

\bibitem{Hatakeyama1997}
N.~Hatakeyama and T.~Kambe.
\newblock Statistical laws of random strained vortices in turbulence.
\newblock {\em Phys. Rev. Lett.}, 79:1257--1260, 1997.

\bibitem{Isett2018}
P.~Isett.
\newblock A proof of {O}nsager's conjecture.
\newblock {\em Ann. Math.}, 188:871--963, 2018.

\bibitem{Kambe2000}
T.~Kambe and N.~Hatakeyama.
\newblock Statistical laws and vortex structures in fully developed turbulence.
\newblock {\em Fluid Dyn. Res.}, 27:247--267, 2000.

\bibitem{Kaneda2003}
Y.~Kaneda, T.~Ishihara, M.~Yokokawa, K.~Itakura, and A.~Uno.
\newblock Energy dissipation rate and energy spectrum in high resolution direct
  numerical simulations of turbulence in a periodic box.
\newblock {\em Phys. Fluids}, 15:{L21--L24}, 2003.

\bibitem{Kida1990}
S.~Kida and S.~A. Orszag.
\newblock Energy and spectral dynamics in forced compressible turbulence.
\newblock {\em J. Sci. Comput.}, 5:85--125, 1990.

\bibitem{Landau1987}
L.~D. Landau and E.~M. Lifschitz.
\newblock {\em Fluid Mechanics -- Volume 6 of Course in Theoretical Physics}.
\newblock Pergamon Press, Oxford, 1987.

\bibitem{Lee_Lele_Moin_1993}
S.~Lee, S.~K. Lele, and P.~Moin.
\newblock Direct numerical simulation of isotropic turbulence interacting with
  a weak shock wave.
\newblock {\em J. Fluid Mech.}, 251:533--562, 1993.

\bibitem{Aluie2019}
A.~Lees and H.~Aluie.
\newblock Baropycnal work: A mechanism for energy transfer across scales.
\newblock {\em Fluids}, 4:92, 2019.

\bibitem{LELE1992}
S.~K. Lele.
\newblock Compact finite difference schemes with spectral-like resolution.
\newblock {\em J. Comput. Phys.}, 103:16--42, 1992.

\bibitem{Lele1994}
S.~K. Lele.
\newblock Compressibility effects on turbulence.
\newblock {\em Annu. Rev. Fluid Mech.}, 26:211--254, 1994.

\bibitem{Mallat1999}
S.~Mallat.
\newblock {\em A wavelet tour of signal processing}.
\newblock Academic Press, New York, USA, 1999.

\bibitem{Onsager1949}
L.~Onsager.
\newblock Statistical hydrodynamics.
\newblock {\em Nuovo Cimento Suppl.}, 6:279--287, 1949.

\bibitem{Pope2000}
S.~B. Pope.
\newblock {\em Turbulent Flows}.
\newblock Cambridge University Press, Cambridge, UK, 2000.

\bibitem{Pullin1998}
D.~I. Pullin and P.~G. Saffman.
\newblock Vortex dynamics in turbulence.
\newblock {\em Annu. Rev. Fluid Mech.}, 30:31--51, 1998.

\bibitem{Pushenko2024}
V.~Pushenko and J.~Schumacher.
\newblock {Connecting finite-time Lyapunov exponents with supersaturation and
  droplet dynamics in a turbulent bulk flow}.
\newblock {\em Phys. Rev. E}, 109:045101, 2024.

\bibitem{Schumacher2007}
J.~Schumacher, K.~R. Sreenivasan, and V.~Yakhot.
\newblock {Asymptotic exponents from low-Reynolds-number flows}.
\newblock {\em New J. Phys.}, 9:{89}, 2007.

\bibitem{Sreenivasan1984}
K.~R. Sreenivasan.
\newblock On the scaling of the turbulence energy dissipation rate.
\newblock {\em Phys. Fluids}, 27:1048--1051, 1984.

\bibitem{Sreenivasan1998}
K.~R. Sreenivasan.
\newblock {An update on the energy dissipation rate in isotropic turbulence}.
\newblock {\em Phys. Fluids}, 10:528--529, 1998.

\bibitem{Sreenivasan2025}
K.~R. Sreenivasan and J.~Schumacher.
\newblock What is the turbulence problem, and when may we regard it as solved?
\newblock {\em Annu. Rev. Condens. Matter Phys.}, 16:121--142, 2025.

\bibitem{Zinchenko2024}
G.~Zinchenko, V.~Pushenko, and J.~Schumacher.
\newblock {Local precursors to anomalous dissipation in Navier-Stokes
  turbulence: Burgers vortex-type models and simulation analysis}.
\newblock {\em Phys. Rev. Fluids}, 9:114608, 2024.

\end{thebibliography}

\end{document}